\documentclass[useAMS,usenatbib]{mn2e}
\usepackage{times}
\usepackage{psfig}
\usepackage{graphics}
\usepackage{lscape}
\usepackage{epsf}
\usepackage{epsfig}
\usepackage{color}

\newcommand{\zoph}{\mbox{$\zeta\,$Oph}}
\newcommand{\zpup}{\mbox{$\zeta\,$Pup}}
\newcommand{\zori}{\mbox{$\zeta\,$Ori}}
\newcommand{\xper}{\mbox{$\xi\,$Per}}

\newcommand{\vinf}{\mbox{$v_\infty$}}
\newcommand{\mdot}{\mbox{$\dot{M}$}}

    \newcommand{\lsim}{\raisebox{-.4ex}{$\stackrel{<}{\scriptstyle \sim}$}}
\newcommand{\msim}{\raisebox{-.4ex}{$\stackrel{>}{\scriptstyle \sim}$}}

\def\langl{\mathopen{<}}
\def\rangl{\mathopen{>}}

\def\changed{}

\title{High resolution X-ray spectroscopy of bright O type stars}

\author[L.\,M. Oskinova, A. Feldmeier, W.-R. Hamann]
{L. M. Oskinova,\thanks{E-mail: lida@astro.physik.uni-potsdam.de}
A.~Feldmeier, W.-R.~Hamann\\ 
Astrophysik, Universit{\" a}t Potsdam, Am Neuen Palais 10,  
Potsdam 14469, Germany}
    
\date{Accepted . Received ; in original changedorm }
 
\pagerange{\pageref{firstpage}--\pageref{lastpage}} \pubyear{2006}

\begin{document}

\label{firstpage}

\maketitle

\begin{abstract} 
{\changed Archival X-ray spectra of the four prominent single,
non-magnetic O stars \zpup,\,\zori,\,\xper,\,\zoph, obtained in high
resolution with Chandra {\sc hetgs/meg} have been studied. The
resolved X-ray emission line profiles provide information about the
shocked, hot gas which emits the X-radiation, and about the bulk of
comparably cool stellar wind material which partly absorbs this
radiation.  In the present paper we synthesize X-ray line profiles
with a model of a clumpy stellar wind.  We find that the geometrical
shape of the wind inhomogeneities is important: better agreement with
the observations can be achieved with radially compressed clumps than
with spherical clumps.  The parameters of the model, i.e. chemical
abundances, stellar radius, mass-loss rate and terminal wind velocity,
are taken from existing analyses of UV and optical spectra of the
program stars. On this basis we also calculate the continuum
absorption coefficient of the cool-wind material, using the Potsdam
Wolf-Rayet ({\sc PoWR}) model atmosphere code. The radial location of X-ray
emitting gas is restricted from analyzing the {\em fir} line ratios of
helium-like ions. The only remaining free parameter of our model is
the typical distance between the clumps; here we assume that at any
point in the wind there is one clump passing by per one dynamical
timescale of the wind. The total emission in a model line is scaled to
the observation. There is a good agreement between synthetic and
observed line profiles.  We conclude that the X-ray emission line
profiles in O stars can be explained by hot plasma embedded in a cool
wind which is highly clumped in the form of radially compressed shell
fragments.}
\end{abstract}

\begin{keywords}
X-rays:stars, stars:individual:\,\zpup,\,\zori,\,\xper,\,\zoph
\end{keywords}


\section{Introduction}

Young and massive O-type stars possess strong stellar winds. The winds
are fast,  with typical velocities up to 2500 km/s, and dense, with
mass-loss rates   $\dot{M} \msim 10^{-7} M_{\odot}/{\rm yr}$. The
driving mechanism for the mass-loss  from OB stars has been identified
with radiation pressure on spectral lines  \citep{CAK75}.  However, it
was pointed out early \citep{Lucy70} and later  further investigated
\citep{OR84} that the stationary solution for a line-driven  wind is
unstable; small perturbation grow quickly and result in strong shocks
giving rise to the production of hot gas and emitting of X-rays.

X-rays from hot OB type stars were discovered by the Einstein X-ray
observatory \citep{har79,sew79}. A large number of OB type stars was
detected by the Rosat All Sky Survey (Bergh{\"o}fer, Schmitt 
\& Cassinelli 1996), providing firm
evidence that the X-ray emission is intrinsic for O star winds. The
low resolution Rosat spectrum of the O-type supergiant \zpup\ was
modeled by \citet{hil93}. In order to explain the apparent weakness of
absorption in the spectrum, \citet{hil93} postulated that the X-rays
are emitted far out in the stellar wind, at distances larger that
$100\,R_*$, by an optically thin shell expanding at constant velocity.
The emission from such shell would produce a broad rectangular
box-like line.

A {\changed box-like} line profile is distinctly different from
{\changed a skewed triangular} line expected from radiation arising in
the wind acceleration zone and attenuated in the wind \citep{mcf91}.
\citet{ig01} demonstrated that the asymmetry of the line correlates
with the wind opacity and density. {\changed In a more optically thick
  wind the lines are more blue-shifted.} Since atomic opacity is
wavelength dependent, the lines are expected to differ in shape across
the spectrum, with larger blue-shift {\changed at wavelengths where
  the wind opacity is higher.}

X-ray generation in the wind acceleration zone is predicted in the
model by Feldmeier, Puls \& Pauldrach (1997), who showed from
hydrodynamic simulations that dense cool shells of gas form in deep
wind regions. Collisions of these shells can lead to strong
shocks. X-rays originate from radiatively cooling zones behind the
shock fronts. The hot X-ray emitting plasma is thermal, has a small
filling factor, and is optically thin. The produced X-rays can be
significantly attenuated by the overlying cool stellar wind.  Hence
the blue-shifted lines are expected.

High-resolution X-ray spectra of O-type stars were obtained for the
first time by the Chandra X-ray observatory \citep{wal01}. The
resolved emission line profiles offered a stringent test for the
theory. The analysis of the line ratios from He-like ions in \zori~and
\zpup~ constrained the regions of X-ray formation relatively close to
the stellar surface. The line width was measured to be less than
corresponding to the maximum velocity, confirming {\changed that the
  line emission originates} in the wind acceleration zone.  Line
blue-shifts were also detected, indicating wind attenuation
\citep{cas01}.

However, the data revealed that the line blue-shifts do not change
significantly  for lines at different wavelengths in contrast to what
was expected \citep{wal01,cas01}. Moreover, the line fitting using  the
''standard model'' of smooth homogeneous wind was not able to  to 
reproduce the lines observed in non-magnetic single O stars unless 
with significantly reduced  mass-loss rates 
\citep[Kramer, Cohen \& Owocki 2003,][]{cohen06}.

{\changed There is a good reason to believe that mass-loss rates from O
stars  are need to be revised downwards. A plenitude of observational 
and theoretical evidence indicates the clumped inhomogeneous nature  of
O star winds \citep[][Bouret, Lanz \& Hillier 2005, Fullerton, 
Massa \& Prinja 2006]{Rep04}. The empirical mass-loss rates of WR stars 
are generally reduced by factors of a few in a clumped wind
\citep{hk98},  but it is still not known by how much exactly in case of
O stars  \citep{puls06}. 

Radiative transfer in a stochastic medium is distinctly different from
the transfer in a smooth medium \citep{pom91}, especially when the
clumps are not necessarily optically thin. Therefore, it is
inconsistent to apply the smooth-wind formalism in order to model
spectral lines in a wind where the mass-loss rate estimates are based
on the assumption of wind clumping.

The transport of X-rays in a clumped stellar wind was studied
analytically and numerically by Feldmeier, Oskinova \& Hamann (2003)
and Oskinova, Feldmeier \& Hamann (2004),
respectively.  Motivated by the prediction of hydrodynamic simulations
that the cool-wind clumps are radially compressed, we considered the
case of anisotropic wind opacity, where the optical thickness of a
clump depends on its orientation.  \citet{feld03} showed and
explained that the assumption of flattened wind clumps
leads to nearly symmetric and blueshifted line shapes.  A
corresponding 2D stochastic wind model developed in \citet{osk04}
demonstrated that clumping effectively reduces the wind opacity and its
wavelength dependence. The latter effect leads to a similarity of the
shapes of all X-ray lines. Recently, \citet{ow06} used a similar
model, but restricted to the specific case of isotropic clumps. In the
latter model a symmetric line profile results only when the mean
separation between optically thick clumps is quite large, of the order
of a few stellar radii. 

The main objective of the present paper is to compare, for the first
time, the  model lines from clumped winds with the lines observed in the
X-ray spectra  of the prominent single non-magnetic O stars
\zpup,\,\xper,\,\zori, and \zoph.

Our basic strategy is to adopt the best available parameters for the
stars and  their winds, compute on their basis the model line profiles,
and compare  them with the  observation. Thus we do not infer the model
parameters  from the line fitting, but instead answer the question
whether the  observed lines can be described in the framework of the
X-ray  transport in stochastic media.

We utilize  the most recent empirical mass-loss rates of our sample
stars,  and evaluate the mass-absorption coefficient in the wind using
state-of-the-art  non-LTE atmosphere code. Observed X-ray spectra are
analyzed to infer the  zone of X-ray emission and the wind velocity
field. The only remaining  free parameter of our model is the average
time interval between subsequent  clumps passing the same point in the
stellar wind. For this parameter we adopt the wind flow time,
$R_*/\vinf$.  

With {\em all} model parameters defined, we model each line and compare 
it with the observation. }

%
\begin{figure*}
\epsfig{figure=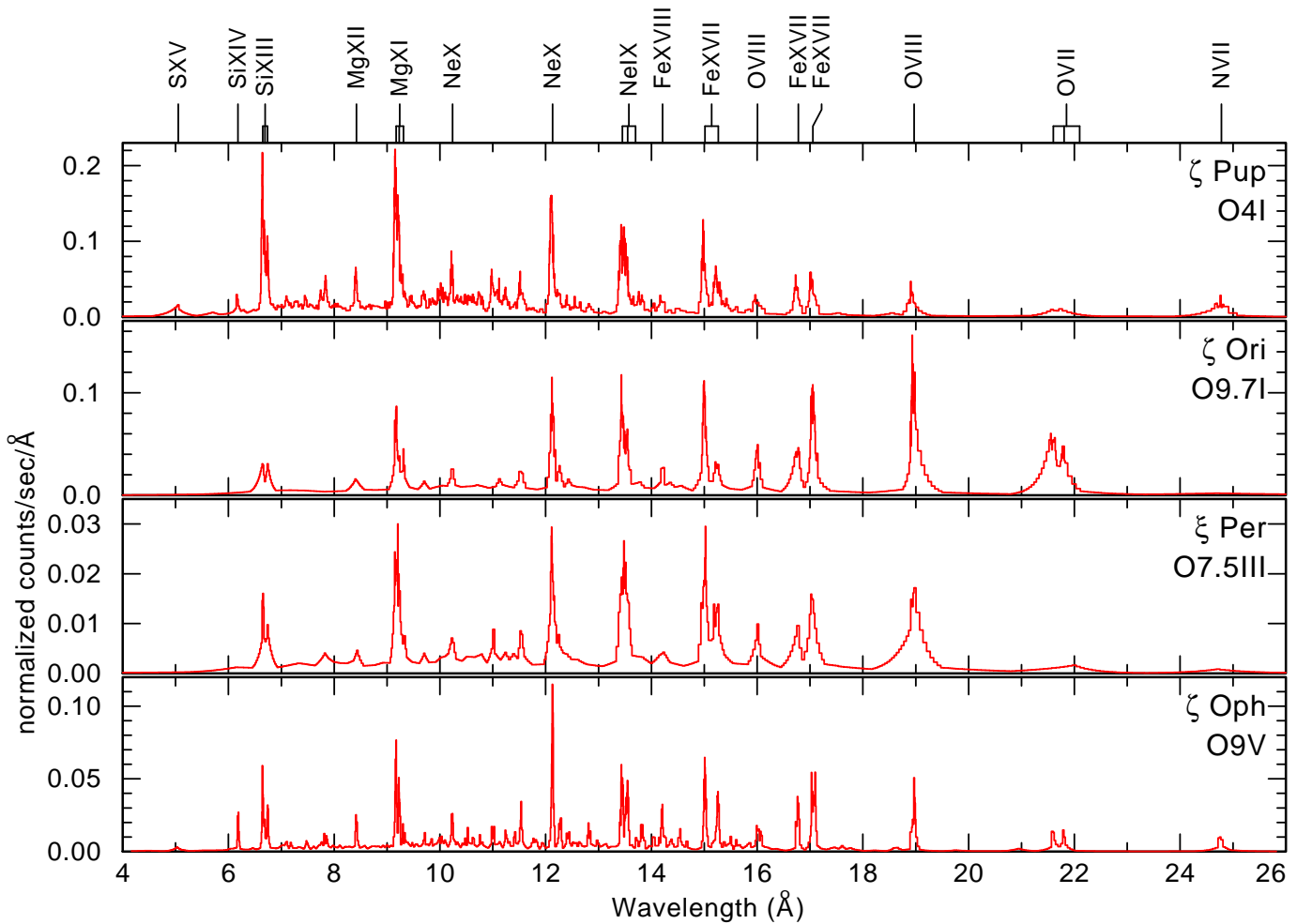, height=18cm, angle=-90}
\caption{De-reddened {\sc hetgs/meg}$\pm1$ spectra of prominent O stars. }
\label{fig:der} 
\end{figure*} 
%
%

The paper is organized as follows. Observational evidences of strong
wind clumping  are discussed in Section~2. The observations and
broad-band spectral properties of  our sample stars are presented in
Section~3. The  analysis of line ratios of He-like  ions, conducted in
order to estimate  the location and temperature of the X-ray emitting
plasma, is presented in  Section~4. Section~5 describes the stellar
atmosphere model, and Section~6 briefly summarizes our radiative
transfer technique for X-rays in a  clumped stellar wind. The
comparison of modeled and observed line profiles  is done in Section~7,
and a discussion follows in Section~8. Conclusions are presented in
Section~9.      

\section{Clumped winds of early-type stars}

%
\begin{table}  
\begin{center}   
\caption{Chandra ACIS-S HETGS observations of prominent O-type stars.}
\vspace{1em} 
\renewcommand{\arraystretch}{1.2}  
\begin{tabular}[h]{lcccc}  \hline
\hline Name  & ObsID & $\log(N_{\rm H})$ &  Flux$^1$  & exposure  \\ 
&       & [cm$^{-2}$]       &[erg/s/cm$^2$]  & [ksec] \\ \hline 
$\zeta$~Pup& 640  &20.00 & $9.0\times 10^{-12}$ & 67  \\  
$\zeta$~Ori& 610  &20.48 & $6.0\times 10^{-12}$ & 59  \\  
$\zeta$~Ori& 1524 &20.48 & $6.0\times 10^{-12}$ & 13  \\  
$\xi$~Per  & 2450 &21.06 & $1.2\times 10^{-12}$ & 160 \\  
$\zeta$~Oph& 4367 &20.78 & $3.1\times 10^{-12}$ & 48  \\ 
$\zeta$~Oph& 2571 &20.78 & $3.1\times 10^{-12}$ & 35  \\ \hline
\hline \\  
\noalign{\smallskip}
\multicolumn{5}{l}{$^1$~Flux not corrected for interstellar absorption}
\end{tabular}  
\label{tab:obs}  
\end{center}  
\end{table} 
%
%

There is mounting observational evidence of strong inhomogeneity of
stellar winds.  Stochastic variable structures in the  He\,{\sc
ii}~4686\,\AA~emission line in \zpup~were revealed by Eversberg, 
L{\`e}pine \& Moffat (1998), and 
explained as an excess emission from the  wind clumps. \citet{mar05} 
investigated the line-profile variability of  H$\alpha$ for a large
sample  of O-type supergiants. They concluded that the properties of
the H$\alpha$  variability can be explained by a wind model consisting
of coherent or  broken shells. \citet{bou05} conducted a quantitative 
analysis of the  far-ultraviolet spectrum  of two Galactic O stars
using the stellar  atmosphere  code {\sc cmfgen}. They have shown that
homogeneous wind models  could not match the observed profiles of
O\,{\sc v} and N\,{\sc iv} and  the phosphorus abundance.  However, the
clumped wind models match well all these lines and are consistent  with
the H$\alpha$ data. This study provided strong evidence of wind 
clumping in O stars starting  just above the sonic point. A reduction
of mass-loss rates by a factor of at least 3 compared to the
homogeneous wind model was suggested. A similar  result was obtained
from fitting stellar wind profiles of the P\,{\sc v} resonance doublet
by \citet{ful06}. A sample of 40 Galactic O stars was studied,  and
the  conclusion was drawn that the mass-loss rates shall be reduced by
up to one order of magnitude, depending on the actual fraction of
P\,{\sc v} in the  wind. The discordance was attributed to the strong
clumping of the stellar  winds. 

X-ray observations of massive stars allow to probe the column  density
of the wind. An analysis of the X-ray emission from the O stars 
$\delta$~Ori \citep{mil02} and $\zeta$~Pup \citep{kr03} have  shown
that the attenuation by the stellar wind is significantly smaller than 
expected from standard homogeneous wind models. 

Similar conclusions are reached from the analysis of X-rays from
colliding wind  binaries (CWB). Such systems consist of two massive
early-type stars. The copious  X-ray emission is produced in the wind
collision zone.  At certain orbital phases the X-ray emission from the
colliding wind zone travel towards the observer through the bulk  of
the stellar wind of one companion.  Deriving the absorbing column
density from  X-ray spectroscopy constrains the mass-loss rates. An
analysis of XMM-Newton  observations of the massive binary
$\gamma^2$\,Vel was presented by \citet{sch04}.  They showed that the
observed attenuation of X-rays is much weaker than expected  from
smooth stellar wind models. To reconcile theory with observation,
\citet{sch04} suggest that the wind is strongly clumped. Similar
conclusions were  reached from  Chandra and RXTE observations of
WR\,140 \citep{pol05}. Alike  $\gamma^2$\,Vel, the column density
expected from the stellar atmosphere models  that account for clumping
in first approximation only is a factor of four  higher than the column
density inferred from the X-ray spectrum analysis. 

%
\begin{table*} 
 \begin{center}  
\caption{Stellar wind parameters }
    \renewcommand{\arraystretch}{1.2} 
    \begin{tabular}[h]{lcccccccc} 
      \hline \hline
Name  & Sp.Type & $R_*$ & $T_{\rm eff}$ & $v_\infty$ & $T_{\rm fl}$ & 
$\dot{M}$ &  clumping$ ^{\rm a}$ & ref \\
           &         & [$R_\odot$] &  [kK]           &[km/s] & [hr] &  
[$M_\odot\,{\rm yr}^{-1}$]& &  \\ \hline 
$\zeta$~Pup& O4I     & 18.6 & 39 & 2250& 1.6 &$4.2\times 10^{-6}$ & yes & 1\\ 
$\zeta$~Ori& O9.5I   & 31   & 32 & 2100& 2.9 &$2.5\times 10^{-6}$ & no & 2\\ 
$\xi$~Per  & O7.5III & 24.2 & 35 & 2450& 1.9 &$1.2\times 10^{-6}$ & yes & 1\\ 
$\zeta$~Oph& O9V     & 8.9  & 32 & 1550& 1.1 &$\lsim 1.8\times 10^{-7}$ & 
no & 3\\
      \hline \hline \\ 
\noalign{\smallskip}
\multicolumn{7}{l}{$ ^{\rm a}$ ``yes'' -- clumping was taking into account 
for mass-loss estimates; ``no'' -- mass-loss is not corrected for the 
clumping}\\
\multicolumn{7}{l}{1)~\citet{puls06}, 2)~\citet{lam93}, 
3)~\citet{Rep04} }  
      \end{tabular} 
    \label{tab:wind} 
  \end{center} 
\end{table*} 
%

Spectacular evidence of wind clumping comes from X-ray spectroscopy of
high-mass  X-ray binaries (HMXB). In some of these systems a neutron
star is on a close  orbit deeply inside the stellar wind of an OB star.
The X-ray emission with  a power-law spectrum results from Bondi-Hoyle
accretion of the  stellar wind onto the NS. These X-rays photoionize
the stellar wind. The resulting  X-ray spectrum shows a large variety
of emission features, including lines from  H-like and He-like ions and
a number of fluorescent emission lines. \citet{sako03}  reviewed
spectroscopic results obtained by X-ray observatories for several
wind-fed  HMXBs. They conclude that the observed spectra can be
explained only as originating  in a clumped stellar wind where cool
dense clumps are embedded in rarefied   photoionized gas. \citet{vdm05}
studied  the X-ray light curve and spectra of  4U~1700-37. They showed
that the feeding of the neutron star by a strongly clumped  stellar
wind is consistent with the observed stochastic variability.

These observational findings appear to be consistent with the stellar
wind theory.  The hydrodynamic modeling of \citet{AF97} (see also a 
pseudo 2D simulation of \citealt{des03}) predicts that
the stellar winds are  strongly inhomogeneous starting from close to
the core, and with large density,  velocity and temperature variations
due to the  de-shadowing instability. This  instability leads to strong
gas compression resulting in dense cool shell fragments  (clumps). The
space between fragments is essentially void, but at the outer side of 
the dense shells exist extended gas reservoirs. Small gas cloudlets are
ablated from  these reservoirs, and accelerated to high speed by
radiation pressure. Propagating  through an almost perfect vacuum, they
catch up with the next outer shell and ram  into it. In this collision,
the gas parcels are heated and emit thermal X-rays.  The X-ray emission
ceases when the wind reaches its terminal speed. In contrast,  the cool
fragments are maintained out to large distances. \citet{RO05} studied 
the 1D evolution of instability-generated structures in the winds and
demonstrated  that the winds are inhomogeneous out to distances of
1000\,$R_*$. Thus theory  predicts the existence of two disjunctive
structural wind components -- hot gas parcels  that emit X-rays, and
compressed cool fragments that attenuate this radiation.

\section{High-resolution X-ray spectroscopy of single O stars} 

A growing number of O-type stars has been observed with the
high-resolution spectrographs on board of Chandra and XMM-Newton. The
purpose of this paper is to study the X-ray line profiles, therefore
we selected the brightest stars with well-resolved lines. The {\sc
hetgs/heg} detector on board of Chandra has a spectral resolution
$\Delta\lambda = 0.012$\,\AA , while {\sc hetgs/meg} has
$\Delta\lambda = 0.024$\,\AA. However the effective area of {\sc heg}
is smaller than that of {\sc meg}. The {\sc rgs} on XMM-Newton has
larger effective area but more modest spectral resolution. For
consistency we concentrate here only on the first order co-added {\sc
hetgs/meg} spectra.

\begin{figure}
\epsfxsize=\columnwidth
\centering \mbox{\epsffile{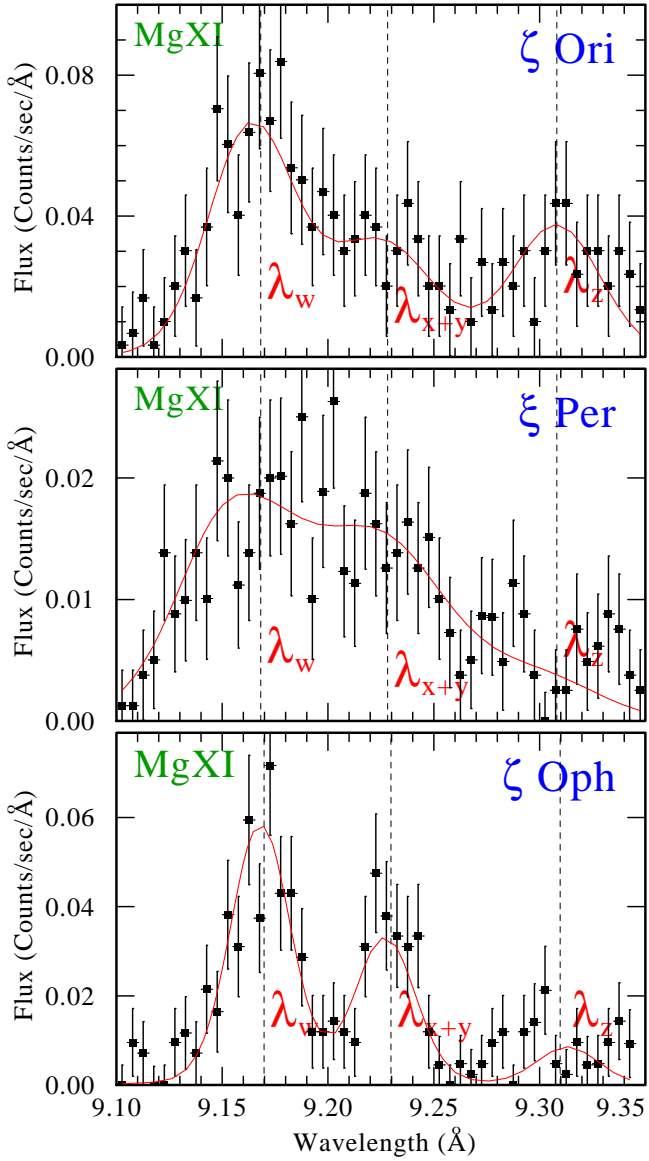}}
\caption{Mg\,{\sc xi} line observed in the spectrum of \zori, 
\xper, and \zoph\ (co-added {\sc meg}$\pm1$). Vertical dashed 
lines indicate rest-frame wavelength: $\lambda_{\rm w}$ -- resonant line,
$\lambda_{\rm x+y}$ -- intercombination line, $\lambda_{\rm z}$ -- 
forbidden line. The rest frame wavelengths in this and all following 
figures are corrected for the  radial velocities taken from 
\citet{Hoog01}. The solid lines are triple-Gaussians fits.}
\label{fig:ne9}
\end{figure} 

We consider only single stars because in a CWB the geometry of the
system can affect the line shapes (Henley, Stevens \& Pittard 2003).
However, there is no way to make sure that an O star is definitely
single. Runaway O stars are distinguished by an almost complete lack
of multiplicity (Hoogerwerf, de Bruijne \& Zeeuw, 2001).  Therefore,
we have selected all three single O-type runaways observed by Chandra
{\sc hetgs} -- $\zeta$~Pup, $\xi$~Per, and $\zeta$~Oph. These stars
are relatively X-ray bright and observed with exposures long enough to
accumulate high quality spectra as shown in Table~\ref{tab:obs} and in
Fig.~\ref{fig:der}. We have also considered a fourth star,
$\zeta$~Ori, albeit it is a known binary \citep{hum00}, because the
initial interpretation of X-ray emission line profiles from this star
caused doubts in the validity of the shock model of X-ray production
\citep{wal01}. The stellar and wind parameters of our sample stars are
compiled in Table~\ref{tab:wind}.

\citet{wal01} and \citet{cohen06} presented a detailed analysis of
Chandra  observations of $\zeta$~Ori. \citet{cas01} and \citet{kr03}
analyzed Chandra  spectra of $\zeta$~Pup. XMM-Newton observations of
$\zeta$~Pup were analyzed  by \citet{kahn01} and \citet{osk06}.
\citet{wal05} reported observations of  $\zeta$~Oph. Chandra
observations of $\xi$~Per were not yet published to our knowledge.  

We retrieved the public archival data of these four stars and extracted
the spectra  using the latest version of the Chandra software and the
calibration data base. The  de-reddened {\sc meg} spectra of the stars
are shown in Fig.~\ref{fig:der}  (see also XMM-Newton spectra of O
stars in Fig.~8 from \citet{paer03}). The emission  lines seen in all
spectra are resolved, with broader lines seen in   the stars with
higher wind velocities. The spectra exhibit emission from H-like and 
He-like ions of low and intermediate Z elements. The Si and Mg lines
are  most prominent in the \zpup~spectrum, while oxygen lines dominate
the spectrum of \zori.  The ratio of nitrogen to oxygen line fluxes is
higher in the runaway stars compared to the $\zeta$~Ori spectrum.
\citet{kahn01} found that in \zpup~the emission  measures derived from
the nitrogen emission lines are at least one order of magnitude  larger
that those of carbon and oxygen. They conclude that the elemental 
abundance ratios of nitrogen to oxygen, as well as nitrogen to carbon
are  considerably higher than  solar indicating that the material has
undergone  CNO processing. This is in  accordance with scenarios where
runaways  have previously been  members of close binary systems and
undergone mass  exchange. 

{\changed The spectral energy distribution (SED) of all stars can be
fitted  using the standard collisional plasma models, e.g.\ {\em apec}.
The spectral  fits indicate the presence of multi-temperature plasma
with temperatures  in the range $ 0.2\,{\rm keV} \lsim kT_{\rm X}\lsim
0.7$\,keV. \citet{kahn01}  reported the detection of continuum in the
XMM-Newton spectrum  of \zpup,  although their fits were inconclusive.
\citet{osk06} analyzed archival  \zpup\ data and concluded that the
lines to continuum ratio is in agreement with the inferred plasma 
temperatures.} 

The emission lines seen in the spectra of O stars cannot be fitted by
means  of the stationary plasma emission line models implemented in the
fitting  software. The standard models are not adequate for the fast
moving stellar  winds, and cannot predict expected line profiles.

\section{Line ratios for helium-like ions}
\label{sec:fir}

\begin{table*}  
\begin{center}  
\caption{Ratios of fluxes$^{\rm a}$ in the components of He-like ions
estimated by fitting observed spectral features.}
\vspace{1em} 
\renewcommand{\arraystretch}{1.2}  
\begin{tabular}[h]{|c|cc|cc|cc|} 
\hline  
Star & \multicolumn{2}{|c|}{Ne\,{\sc ix}} & 
       \multicolumn{2}{|c|}{Mg\,{\sc xi}} & 
       \multicolumn{2}{|c|}{Si\,{\sc xiii}}\\ \hline 
          & R & G & R & G & R & G          \\  
$\zeta$~Pup 
&$0.35\pm 0.12$ & $1.3\pm 0.3$ & $0.36\pm 0.09$ & $1.05\pm 0.15$ &
$0.9\pm 0.2$ & $1.1\pm 0.2$\\ 
$\zeta$~Ori & 
$0.3\pm 0.2$ & $0.9\pm 0.2$ & $0.9\pm 0.4$  & $1.1\pm 0.3$ & 
$1.9\pm 1.6$ & $1.4\pm 0.8$\\ 
$\xi$~Per   & 
$0.1\pm 0.1$ & $1.4\pm 0.5$ & $0.19\pm 0.15$ & $1.17\pm
0.35$ & $2\pm 13 $   & $1\pm 3$\\  
$\zeta$~Oph & 
$1.1\pm 1.1$ & $1.13\pm 0.35$ & $0.3\pm 0.1$  & $1.0\pm 0.3$ & 
$0.9\pm 0.8$ & $0.6\pm 0.3$ 
\\ 
\hline 
\noalign{\smallskip} 
\multicolumn{7}{l}{$^{\rm a}$ Fluxes are from the triple-Gaussian fitting of 
{\sc meg} spectral  lines}  
\end{tabular} 
\label{tab:fir}  
\end{center}  
\end{table*} 
%

\begin{figure}
\epsfxsize=\columnwidth 
\centering \mbox{\epsffile{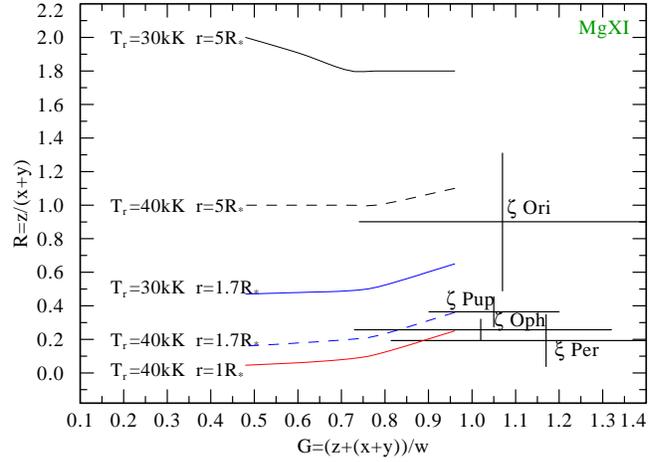}}
\caption{Observed and theoretical $R$ and $G$ ratios for the Mg\,{\sc xi} 
{\em fir} triplet. Observed values are denoted by the star name.
The curves represent the theoretical $R$ and $G$ from \citet{por01} 
for the corresponding  $T_{\rm rad}$ and the radii 
of line formation. Observed values of $G$ are consistent with electron 
temperatures $T_{\rm e}\la 3$\,MK. Observed values of $R$  roughly 
constrain the line emission regions between $\ga 1.5$ and $\la 5\,R_*$ 
for all stars.}
\label{fig:mg11rg} 
\end{figure} 
The He-like ions show characteristic ``{\em fir} triplets'' of a
forbidden ($z$),  an intercombination ($x+y$) and a resonance ($w$)
line. In OB type stars, the {\em fir}  line ratios can be used to
constrain the temperature and location of the X-ray emitting  plasma.
As was shown by \citet{gab69} the ratios $R(n_{\rm e})$ and $G(T_{\rm
e})$ are sensitive to the electron density and to the electron
temperature: 
\begin{equation} 
R(n_{\rm e}, T_{\rm rad})=\frac {z}{x+y} 
\label{eq:rr} 
\end{equation}      
\begin{equation} 
G(T_{\rm e})=\frac {z+(x+y)}{w} 
\label{eq:gr} \end{equation}      
A strong radiation field can lead to a significant depopulation of the 
upper level  of the forbidden line via photo-excitation to the upper
levels of the intercombination lines. This is analogous to the  effect
of electronic  collisional excitation, thus mimicking a high  density.
For the characteristic  densities of O star winds the photo-excitation
is the dominant mechanism  for depopulation. Since the radiation  field
dilutes  with distance  from the stellar surface, the ratio between
forbidden  and intercombination line provides information about the
distance of the X-ray emitting plasma  from the photosphere. However,
this is  possible only when the radiative  temperature $T_{\rm rad}$ of
the radiation field at the wavelength of the   depopulating transition
is known. \citet{por01} performed improved calculations  of $R$ and $G$
line ratios for plasmas in collisional equilibrium and tabulated  them
for a wide range of parameters, including $T_{\rm rad}$. 

Lines of O\,{\sc vii}, Ne\,{\sc ix}, Mg\,{\sc xi}, and Si\,{\sc xiii}
are observed  in all our sample stars, however O\,{\sc vii} is well
resolved only in $\zeta$~Oph.  Therefore, we use in our analysis only
He-like ions of Ne, Mg, and Si. In the sample stars with  large wind
velocities, $\zeta$\,Pup and $\xi$\,Per, the $w$,\,$x+y$, and $z$
components strongly overlap. Hence flux estimates in each separate
component  are ambiguous. The situation is best for $\zeta$~Oph, the
star with the   lowest wind velocity, where all components are well
separated. To obtain  the flux  in the $w$, $x+y$, and $z$ components
for a given ion, we have  fitted all three  components  simultaneously
as Gaussians. Figure\,\ref{fig:ne9}   shows an example of the observed
and fitted lines of a He-like ion. The  derived $R$ and $G$ ratios for
all stars  are listed in Table~\ref{tab:fir}. 

The tables from \citet{por01} were used to constrain the temperature
and location of the X-ray emitting plasma. The ratio $G$ is sensitive
to $T_{\rm e}$,  while it is almost insensitive to $n_{\rm e}$, $T_{\rm
rad}$ and the dilution  factor $W$. In our stars  $G$ is  large. The
derived electron temperatures  are lower than the temperatures
corresponding to the ionisation potential for the  given ion. The
emission measures are highest for the ions with lowest ionisation 
potential. This is in accordance with the  analysis of differential
emission measure in $\zeta$~Pup and $\zeta$~Ori conducted by
\citet{woj05}. 

For electron densities less than $\sim 10^{12}$\,cm$^{-3}$ and in the
presence of a strong UV field, the ratio $R$ (forbidden to
intercombination lines) is not sensitive to the density but depends
instead on the electron temperature $T_{\rm e}$, radiative temperature
$T_{\rm rad}$, and the dilution factor $W$. The electron temperature
is constrained by $G$.  The radiative temperature of the radiation
field at the wavelength of interest (see Table\,3 in \citet{por01})
can be determined from stellar atmosphere models. We employ {\sc PoWR}
model to calculate the radiation temperature as function of wavelength
for different depth points in the wind. The radiative temperature at
$\lambda_{{\rm f}\rightarrow {\rm i}}\,864\,$\AA\,(Si\,{\sc xiii}) as
$\la 90$\,per cent of the effective temperature. For the lighter ions
$\lambda_{{\rm f}\rightarrow {\rm i}}\,1033\,$\AA\,(Mg\,{\sc xi}),
$\lambda_{{\rm f}\rightarrow {\rm i}}\,1270\,$\AA\,(Ne\,{\sc ix}), the
wavelength of interest is on the red side of the Lyman jump in the
stellar SED, and the radiative temperature is higher, roughly equal to
the effective temperature.

\begin{figure}
\epsfxsize=\columnwidth 
\centering \mbox{\epsffile{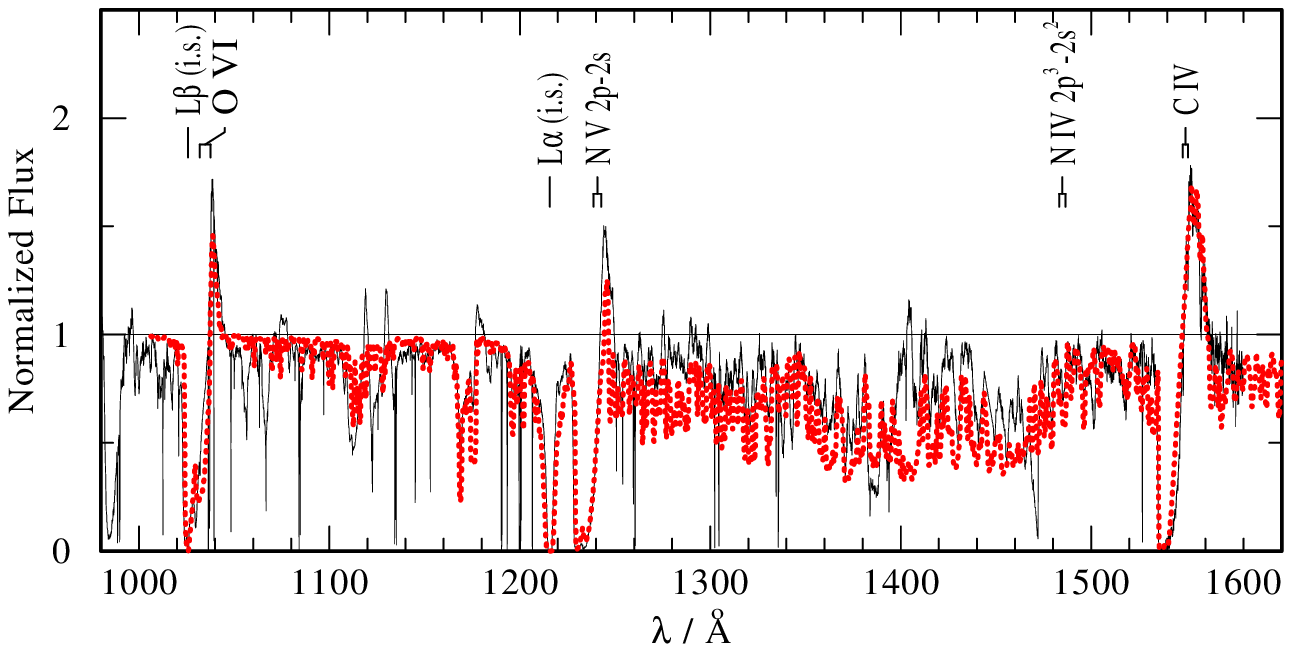}}
\caption{The EUV spectrum of \zpup, observed with FUSE (thin line),
compared to a {\sc PoWR} model spectrum (dotted). The resonance doublets of
C\,{\sc iv}, N\,{\sc v} and O\,{\sc vi} are well reproduced, as well as
the ``forest'' of iron-group lines. The O\,{\sc vi} doublet can only be
fitted with models when assuming that a diffuse X-ray field causes
additional ionisation. }
\label{fig:spmod} 
\end{figure} 

With $T_{\rm rad}$ and $T_{\rm e}$ constrained, the ratio $R$ can be
used  to estimate the dilution factor $W$ and, therefore, the radial
distance $r$ from the stellar photosphere at which the X-rays are
produced.  The ratios $R$ and  $G$ and corresponding error bars
measured in Mg\,{\sc xi}  for our sample stars are  shown in the $R-G$
diagram in Fig.\,\ref{fig:mg11rg}. Theoretical values of  $R$ and $G$, 
calculated by \citet{por01} for  $T_{\rm rad}=30$\,kK and 
$T_{\rm rad}=40$\,kK, and for different distances $r$, are also plotted in 
Fig.\,\ref{fig:mg11rg}. From
comparison between observed  and theoretical values of $R$ and $G$ we 
conclude that the Mg\,{\sc xi} line is emitted by  gas located between
$1.5\,R_*$ and $5\,R_*$ {\changed in all stars. It should be noted, however, 
that the grids presented in \citet{por01} are quite coarse. Moreover, the
observed lines, especially in  \xper, are very noisy. 

We constructed  similar diagrams also for  Ne\,{\sc ix} and Si\,{\sc xiii}. 
The latter line is sufficiently well resolved only in $\zeta$~Pup 
and $\zeta$~Oph. Comparison of $R-G$ diagrams for the different He-like 
ions in our sample stars allows to conclude that the X-ray 
emission originates not closer to the stellar photosphere than 
$\approx 1.5\,R_*$ and not further than $\approx 10\,R_*$, i.e. 
{\changed in the wind acceleration region for all stars of our sample. }   

\begin{figure}
\epsfxsize=\columnwidth
\centering \mbox{\epsffile{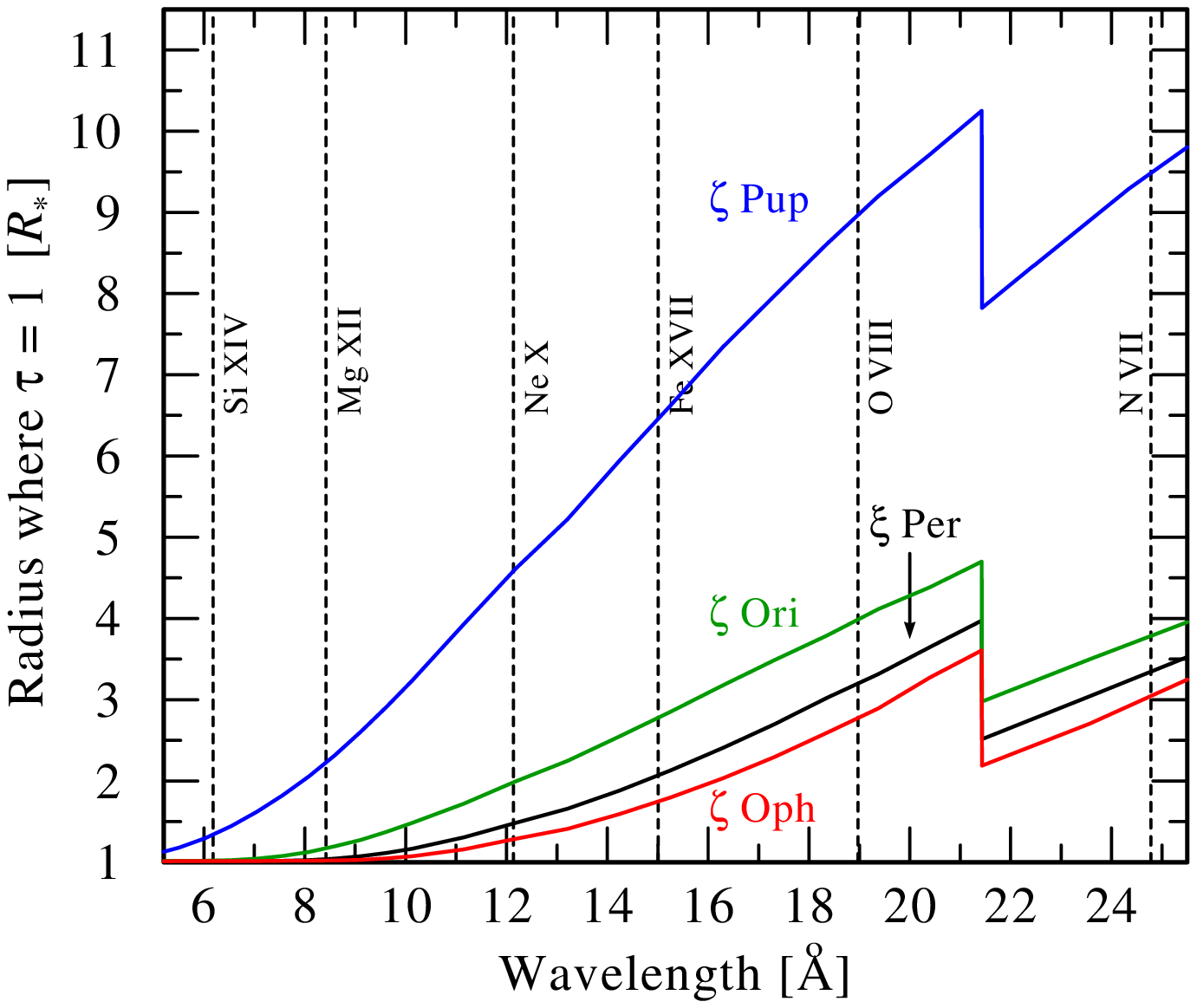}}
\caption{The radius were the radial optical depth of the wind becomes
unity in dependence on the wavelength in the Chandra {\sc hetgs/meg}
range. The calculations were done using the {\sc PoWR} stellar atmosphere
code (see text) with stellar parameters from Table\,\ref{tab:wind}.
The prominent edge at $\lambda\,21.5$\,\AA~is due to oxygen. The
vertical dashed lines correspond to the wavelengths of the
studied lines (as indicated). }
\label{fig:tau1}
\end{figure} 

\section{Mass-loss and stellar atmosphere models of sample stars}

Full non-LTE stellar atmosphere models of O star winds are necessary
to  infer parameters of general, cool stellar wind. Recently,
\citet{puls06} presented an up-to-date review of the effects of clumping
on \mdot \ determinations  in  O star winds. They analyzed H$\alpha$,
IR, and radio fluxes of \zpup\ and  \xper\ (within a sample of 19
Galactic O stars) using non-LTE models which  account for
line-blanketing and clumping. The derived stellar parameters are 
listed in Table~\ref{tab:wind}. The mass-loss rates are identical for
H$_\alpha$,  IR, and radio determinations and constitute upper limits,
since the clumping  properties of the wind are still unknown (see
\citet{puls06} for details). The  stellar parameters for \zoph\ listed
in Table~\ref{tab:wind} are from  \citet{Rep04}. The mass-loss is based
on the analysis of photospheric lines  from H and He and shall be
considered as an upper limit. This is also the case  for parameters of
\zori\ in Table~\ref{tab:wind} \citep{lam99}.  

The aim of this paper is to model the attenuation of X-rays in a
clumped  wind, where the photons are absorbed only in clumps and can
escape freely  between them. The total mass of the wind is conserved
and distributed  stochastically, in the form of clumps. The optical
depth of clumps for the X-rays  is determined by the wind density and
the mass absorption coefficient. In our  clumped model the total {\em
radial} optical depth of the wind is absolutely  the same as for a
homogeneous wind of the same mass-loss rate. 

We consider {\em continuum} absorption of X-rays in clumped cool wind
caused  predominantly by bound-free and K-shell ionisation processes.
Introducing  Cartesian coordinates, the $z$ axis points towards the 
observer and the  impact parameter $p$ stands perpendicular to the
latter.  Alternatively, a  point $(p,z)$ is specified by spherical
coordinates, i.e.\ the radius  $r=\sqrt{p^2+z^2}$ and the angle
$\vartheta=\cos^{-1}\mu$ between  the line  of sight and the radial
vector. Hence $\mu = z/r$. It is assumed that the  time-averaged
stellar wind is spherically symmetric and expands with a
radius-dependent velocity  
\begin{equation}   
v(r)= v_\infty \left(1-\frac{r_0}{r}\right)^\beta,  
\label{eq:vr}   
\end{equation}  
where $v_\infty$ is the terminal velocity, and $r_0$ is chosen such
that  $v(r\!=\!R_\ast)=0.01\,v_\infty$. $R_\ast$ is the stellar
(photospheric) radius. For a given mass loss rate \mdot, the continuity
equation defines the density  stratification via $\rho(r)=\mdot/(4\pi
r^2 v(r))$.  

The wind can be specified by its total radial  optical depth,
\begin{equation} 
\tau_\ast =
\int_{R_*}^{\infty} \kappa_{\nu} \rho(r)\ {\rm d}r\ , 
\label{eq:tast} 
\end{equation}
{\changed where $\kappa_\nu$ is the mass absorption coefficient, which
describes  the continuum opacity of the bulk of ``cool'' wind material
at the frequency of the considered X-ray line.  

In order to provide realistic values for this $\kappa_\nu$, we employ
the Potsdam Wolf-Rayet ({\sc PoWR}) model atmosphere code \citep{powr}. For
each star we calculate a model with the para\-meters from 
Table~\ref{tab:wind}, and the appropriate chemical composition from the
references given in Table~\ref{tab:wind}. For instance, enhanced
nitrogen and depleted carbon and oxygen \citep{lam99} are taken for the
\zpup\ model. Figure~\ref{fig:spmod} shows that our model spectrum of
\zpup\ agrees well with the observed EUV spectrum. 

The O\,{\sc vi} resonance doublet (see Fig.~\ref{fig:spmod}) can only
be reproduced by models if there is a diffuse X-ray field causing some
``superionisation'', as already recognized by \citet{cas79}. In our
model we adopt thermal bremsstrahlung emission from a hot embedded
plasma component of $3\times 10^6$\,K and adjust the filling factor
such that the level of emerging X-rays corresponds roughly to the
observation.  

In fact, $\kappa_\nu$ chiefly depends on the chemical composition, but 
not much on other details of the model, nor on the radial coordinate.
Table\,\ref{tab:kappa} gives $\kappa_\nu$ at  the wavelength position
of the studied X-ray emission lines, for each star of our sample. In
order to visualize how opaque or transparent the stellar winds are for
X-ray radiation, we plot at which radii the radial optical depth
becomes unity (Fig.\,\ref{fig:tau1}), for the wavelength range under
consideration. At a few stellar radii, i.e.\ in the acceleration zone
of the stellar wind where X-ray producing shocks should be located, the
wind of \zpup\ is optically thick for all considered X-ray lines
(except perhaps Si\,{\sc xiv}). In \xper\ and \zori, only the softer of
the X-ray lines should encounter optically thick continuum absorption.
The outer wind of \zoph\ is almost transparent for the X-ray line
photons. }

\section{Transfer of X-rays in a clumped stellar wind}

Wind inhomogeneity alters the radiative transfer significantly. We 
have studied  the effects of wind fragmentation on X-ray line formation
analytically in the  limit of infinitely many stochastically
distributed fragments \citep{feld03},  and numerically for a finite
number of fragments within an accelerating stellar wind \citep{osk04}.
The latter work describes in detail the model and the code  used in the
present paper. Here we briefly summarize the description  of
fragmented-wind opacity in order to introduce the parameters that will 
be used in the next section  to model observed X-ray lines.  

For a homogeneous wind, the optical depth  $\tau_{\rm h}(p,z)$ between
a point $(p,z)$ and the observer is
\begin{equation}
\tau_{\rm h}(p,z) = \frac{\kappa_\nu \mdot}{4\pi}
\int_{z}^{\infty}\frac{{\rm d}z'}{{r'^2 v(r')}}\ .
\label{eq:tauhz}
\end{equation}
with $r' = \sqrt{p^2 + z'^2}$.

\begin{figure}
\epsfxsize=\columnwidth 
\centering 
\mbox{\epsffile{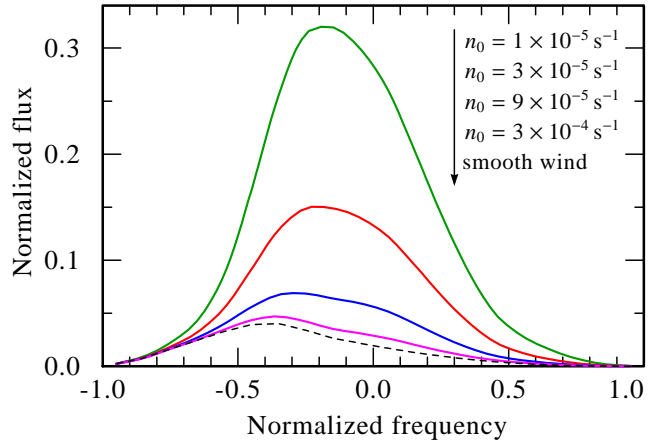}}
\caption{Solid lines: model line profiles of O\,{\sc viii} in the 
spectrum of $\zeta$~Ori for different fragmentation frequency, $n_0$, 
as indicated.  Dashed line: smooth  wind model line. 
The line emission region is assumed between 1.5$R_*$ and 9$R_*$, 
velocity law is with $\beta=1$. Frequency is measured  relative to the 
line center and in Doppler units referring to the terminal  wind velocity 
$v_\infty$. {\changed Flux is normalized to the line flux in the 
unabsorbed case.  All model lines in this paper are convolved with a 
Gaussian according to the {\sc hetgs/meg} spectral resolution. }
\label{fig:linf}}  
\end{figure} 

\begin{figure}
\epsfxsize=\columnwidth 
\centering 
\mbox{\epsffile{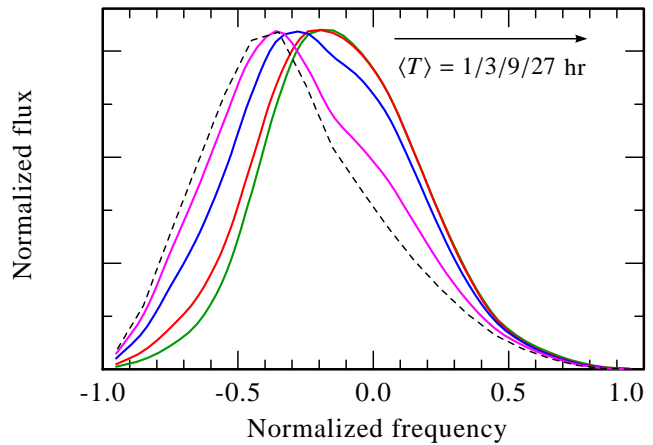}}
\caption{The same as in Fig.\,\ref{fig:linf} but lines are normalized to 
the same maximum. The average time between subsequent clump passage, 
$\langl T \rangl = 1/n_0$, is indicated in the upper right corner.  
\label{fig:linn}}  
\end{figure} 

We assume that close to the stellar core at distances smaller than 
$R_{\rm min}\la 1.5R_*$ the wind is homogeneous and no X-ray  production
is taking place. The wind instabilities start to develop at the 
distances larger than $R_{\rm min}$. These instabilities result in the 
heating of some small fraction of the material to  X-ray emitting 
temperatures. The general stellar wind remains cool and is compressed
into shell  fragments. While general wind fragmentation persists till
large distances  $R_{\rm max}\approx 300 R_*$, the X-ray emission is
ceased  at the  distance $R_{\rm emis}\ga 5 R_*$.  {\changed We assume
that the wind is homogeneous at distances  $r\msim\,R_{\rm max}$ and
take into account the absorption of X-rays by  this part of the wind as
well.}

The hot material ($T_{\rm X}\sim 1$\,MK) emitting X-rays is permeated
with the  cool ($T_{\rm w}\sim 10$\,kK) wind fragments (i.e. clumps),
which attenuate the X-ray emission. Both the hot and the cool gas
component move outwards with  the same velocity $v(r)$ {\changed  and
distributed stochastically. 

We assume for simplicity that all parcels of gas which emit X-rays have
the  same temperature $T_{\rm X}$ and, on average, the same mass. The
line  emission is powered by collisional excitation and therefore
scales with the  square of the density. We assume that the hot gas
parcels expand according  to the continuity equation when propagating
in radial direction.  Hence,  the emission of each hot parcel is
$L_0/(r^2 v(r))$, with a  constant $L_0$ determined by the line
emissivity. 

The mass and the solid angle of each cool wind fragment are also 
preserved.}  In order that the time-averaged mass-flux \mdot\, of
shells resembles that of  a stationary, homogeneous wind, the radial
number density of shells must scale  with $1/v(r)$:
\begin{equation}
n(r) = \frac{n_0}{v(r)}\ .
\label{eq:def_n(r)}
\end{equation}

The parameter $n_0=n(r)v(r)$ defines the number of fragments passing
through some reference radius (e.g. entering or leaving the wind) per
unit of time and is constant due to the mass conservation. It does not
depend on the distance and is therefore convenient model parameter. We
refer to $n_0$ as the {\em fragmentation frequency}.  {\changed The
inverse quantity, $\langl T \rangl$, is the average time interval
between two consequent shells passing the same point in the stellar
wind. When $\langl T \rangl$ is equal to the flow time, 
$T_{\rm fl}=R_*/\vinf$, the average separation between clumps is $1\,R_*$
when $v(r)=\vinf$.  In the wind acceleration region, where the
X-rays are emitted, the radial separation between clumps is smaller.
All model lines presented in this paper are computed assuming 
$n_0=T_{\rm fl}^{-1}$.}

When $n_0$ is specified, the time averaged number of  fragments in
radial  direction is
\begin{equation}
\langle N_{\rm r} \rangle = n_0\int_{R_{\rm min}}^{R_{\rm max}}
\frac{{\rm d}r}{v(r)}.
\label{eq:nre}
\end{equation}

As was discussed in \citet{osk04}, the average {\em radial} optical
depth of a  fragment located at distance  $r$ is
\begin{equation}
\bar{\tau}_j^{\rm rad}=
\frac{\kappa_\nu\dot{M}}{4\pi}\cdot \frac{1}{r^2}\cdot\frac{1}{n_0}.
\label{eq:taujrad}
\end{equation}

We assume here that the cool fragments are compressed radially and
therefore aligned (this model sometimes is referred to as ''Venetian
blinds'').  The optical depth of a flattened fragment $j$ along
line-of-sight depends on its orientation
\begin{equation}
\bar{\tau}_j= \frac{\bar{\tau}_j^{\rm rad}}{|\mu|},
\label{eq:tauj}
\end{equation}
where
\begin{equation}
|{\mu}|=\frac{z_j}{\sqrt{{p^2}+{z_j^2}}}.
\label{eq:muj}
\end{equation}
{\changed Thus, the optical depth of each fragment, and therefore the 
effective wind opacity, is angular dependent.}   As discussed in
\citet{osk04}, the optical depth along the line of sight  between point
$(p,z)$ and the observer in a  stellar wind which consists of  aligned
fragments is
\begin{equation}
\tau(p,z) = n_0\int_{z}^{z_{\rm max}} 
\ (1-{\rm e}^{-\bar{\tau}_j})\,|\mu(r')|\,\frac{{\rm d}z'}{v(r')}+
{\changed \tau_{\rm h}(p,z_{\rm max})},
\label{eq:tint}  
\end{equation}
where {\changed $z_{\rm max}=\sqrt{R_{\rm max}^2-p^2}$, and   ${\rm
d}z'={\rm d}r'/\mu$. In the specific case of isotropic wind  opacity,
which was considered in \citet{ow06},  factor $\mu(r)$  in omitted in
Eq.\,(\ref{eq:tauj}) and Eq.\,(\ref{eq:tint}).}

Let us compare the optical depth in a fragmented wind
(Eq.\,\ref{eq:tint}) with  the optical depth in a homogeneous wind
(Eq.\,\ref{eq:tauhz}). In the limiting  case of optically thin
fragments ($\bar{\tau}_j^{\rm rad}\ll 1$) the exponent  under the
integral in Eq.\,(\ref{eq:tint}) can be expanded. Substituting the
average  optical depth of a fragment defined by Eq.\,(\ref{eq:tauj})  
the optical depth in the thin-fragment limit is the same as in a 
homogeneous wind.

{\changed The dependence of the model line on $n_0$ is demonstrated in 
Figs.\,\ref{fig:linf},\ref{fig:linn}. One can see from
Eq.\,(\ref{eq:taujrad}) that when $n_0$ is small  the fragments are 
opaque for X-rays. In this case the radiation can escape only between  
fragments and the shape of the line profile does not change as long as
the  fragments remain opaque.  With increasing $n_0$ the number of
optically thin fragments increases, changing the line profile. For
sufficiently large $n_0$,   when all clumps are optically thin, the
line profile looks like  from the smooth wind model.}

In the limit of optically thick fragments ($\bar{\tau}_j^{\rm rad}\gg
1$)  the optical depth becomes
\begin{equation}
\tau(p,z) = n_0\int_{r}^{R_{\rm max}} \frac{{\rm d}r'}{v(r')},~~~\tau_j\gg 1\ ,
\label{eq:thick}  
\end{equation}
where the contribution from $\tau_{\rm h}(p,z(R_{\rm max}))$ is omitted
for  clarity. 

The dependence of optical depth on the mass absorption coefficient 
$\kappa_\nu$ and \mdot \ has  disappeared in Eq.\,(\ref{eq:thick}). 
{\changed Thus, in a limiting case of a wind which consists only of
opaque fragments,} the opacity is grey and is  determined by
the  fragmentation frequency $n_0$ and the velocity.

%
\begin{figure*} 
\epsfig{figure=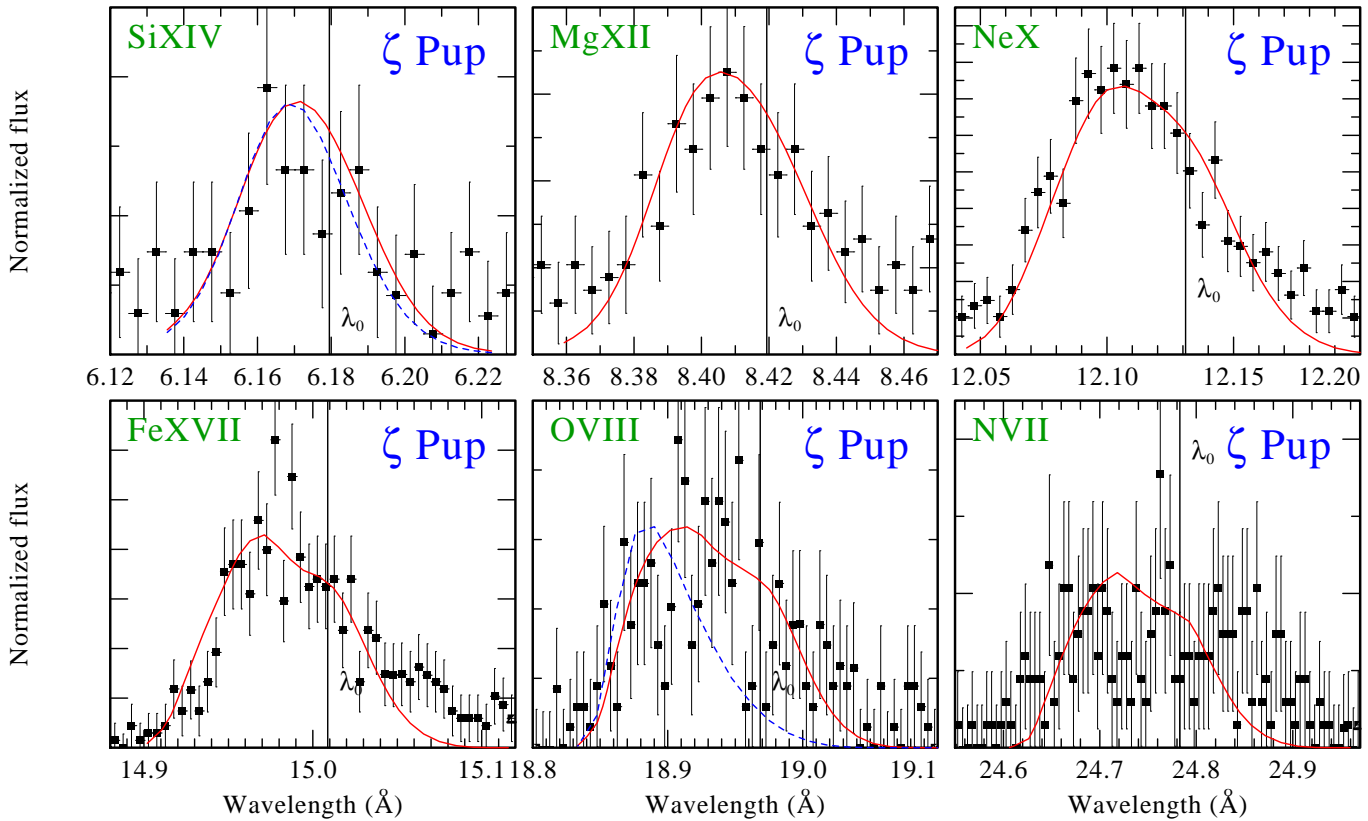, height=18cm, angle=-90} 
\caption{Strongest emission lines observed in \zpup\ 
(co-added {\sc meg}$\pm 1$) and the model lines. The solid lines are 
the model with 
$n_0=1.7\times 10^{-4}~{\rm s^{-1}}$~($\langl T \rangl=1.6$\,hr), 
$\beta=0.9$, and the radiation originating between 1.5$R_*$ and 5$R_*$. 
The dashed line is for the homogeneous model of the same \mdot.  
\label{fig:pupl1}} 
\end{figure*} 
%

In our stochastic numeric wind model the fragments may have different 
optical depths, {\changed and are not necessarily opaque}. As can be
seen  from Eq.\,(\ref{eq:tint}) {\changed even when the optical
thickness of  the fragments is $\sim 1$} the dependence of optical
depth $\tau(p,z)$   on $\kappa_\nu$ is weaker, compared to the
homogeneous  case (Eq.\,\ref{eq:tauhz}).  

Hence, albeit the opacities $\kappa_\nu$ differ significantly depending
on the  wavelength (see Table~\ref{tab:kappa}), the optical depths
$\tau(p,z)$ for  different lines remains similar (more so in the case
of optically thick clumps). Therefore the profiles of X-ray emission
lines are similar across the  spectrum. 

\section{Comparison with observed X-ray line profiles}

\begin{table} 
  \begin{center}  
   \caption{Mass absorption coefficient $\kappa_{\nu}$\,[cm$^{2}$\,g$^{-1}$] 
at the wavelengths of X-ray emission lines from  stellar wind models, 
for the radial range between $1.5R_*$ and $5R_*$}   
    \renewcommand{\arraystretch}{1.2} 
    \begin{tabular}[h]{lccccc} 
      \hline \hline
Line   & Wavelength [\AA] & $\zeta$~Pup & $\zeta$~Ori & $\xi$~Per & $\zeta$~Oph \\
\hline
Si\,{\sc xiv} &  6.18 & 18  & 34  & 11  & 11 \\  
Mg\,{\sc xii} &  8.42 & 22  & 44  & 23  & 23 \\
Ne\,{\sc x}   & 12.14 & 59  & 73  & 55  & 57 \\
Fe\,{\sc xvii}& 15.01 & 94  & 108 & 95  & 95 \\
O\,{\sc viii} & 18.97 & 171 & 177 & 170 & 172 \\
N\,{\sc vii}  & 24.78 & 93  & 80  & 93  & 94 \\   
     \hline \hline \\ 
      \end{tabular} 
    \label{tab:kappa} 
  \end{center} 
\end{table} 

\subsection{Model specification}

{\changed The model is specified by mass-loss rate \mdot, velocity
field  parameters \vinf\ and $\beta$, mass-absorption coefficient
$\kappa_\nu$, and  fragmentation frequency $n_0$, as well as by the 
radii of the line emission zone.  

We use the most recent published  mass-loss rates for our sample stars.
In case  of \zpup\ and \xper\ \mdot\ is corrected for clumping, while 
for \zori\ and \zoph\ the clumping is not accounted for. Velocity
field  parameters are from spectral analyses referenced in
Table\,\ref{tab:wind}. The mass-absorption coefficients are determined
using the {sc PoWR} code. For all our models we adopt the fragmentation
frequency which is equal  to the inverse wind flow time,
$n_0=\vinf/R_*$.  The radii of the line emission zone are constrained
from our analysis of {\em fir} line ratios in Sect.\,4. Thus, no free 
parameter remains and a model is completely specified. } 

%
\begin{figure*}
\epsfig{figure=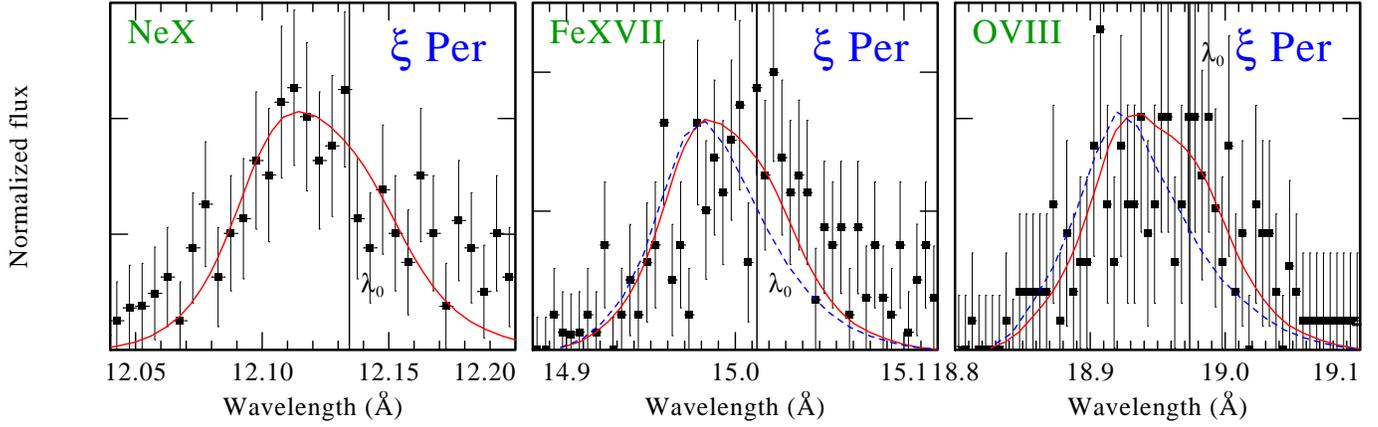, height=18cm, angle=-90}
\caption{Lines of Ne\,{\sc x}, Fe\,{\sc xvii}, and O\,{\sc viii} observed 
in \xper\ (co-added {\sc meg}$\pm 1$) and the
model lines. The solid lines are the model with 
$n_0=1.4\times 10^{-4}~{\rm s^{-1}}$
($\langle T \rangle=1.9$\,hr), $\beta=0.9$, and the radiation originating
between 1.5$R_*$ and 7$R_*$.  Dashed lines are homogeneous wind model.  
\label{fig:xperl}} 
\end{figure*} 

\subsection{X-ray emission lines in \zpup,\,\xper,\,\zori, and \zoph}

{\bf \zpup.}~Figure\,\ref{fig:pupl1} shows the strongest emission
lines observed in \zpup. As can be seen in Fig.\,\ref{fig:tau1}, the
wind of \zpup \ becomes optically thin for X-rays at the Si\,{\sc xiv}
$\lambda 6.18$ \ line just above 2$R_*$. When the wind is optically
thin above the line emission region, the emergent line should be
nearly symmetric relative to the rest wavelength $\lambda_0$. However,
the peak of the Si\,{\sc xiv} line is shifted to the blue, quite
similar to other lines in Fig.\,\ref{fig:pupl1}. This indicates that
the line photons must have originated below the surface of optical
depth unity at $2R_*$. This is in agreement with the estimates based
on the {\em fir} analysis.

\begin{figure*}
\epsfig{figure=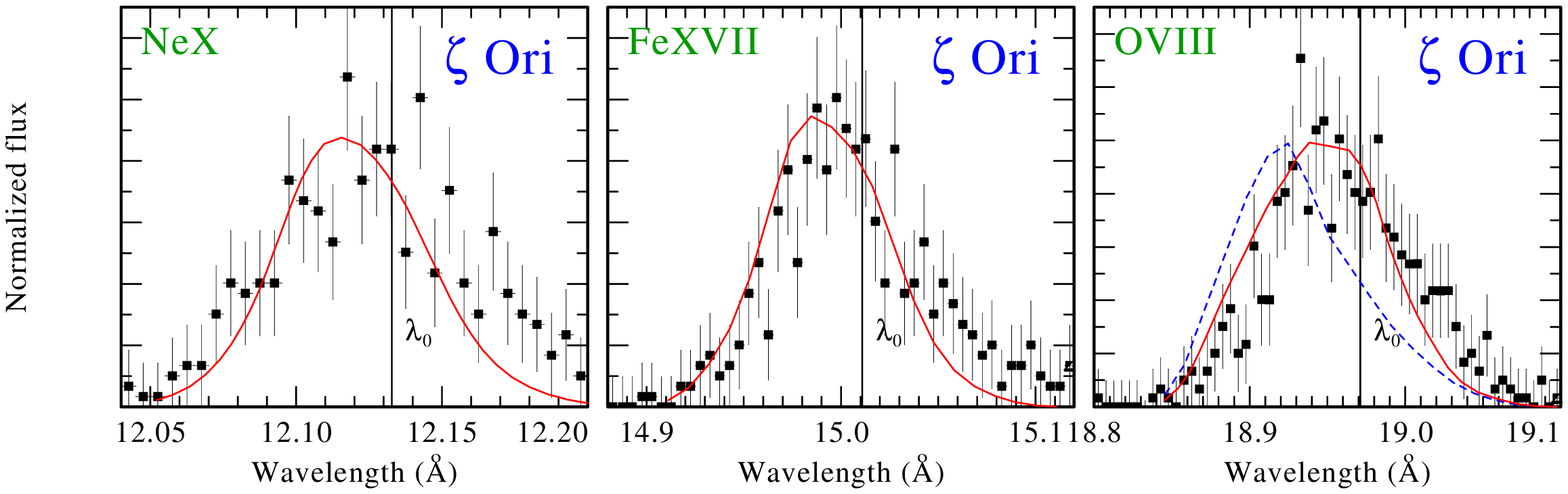, height=18cm, angle=-90}
\caption{Lines of Ne\,{\sc x}, Fe\,{\sc xvii}, and O\,{\sc viii} 
observed in \zori\ (co-added {\sc meg}$\pm 1$) and the
model lines. The solid lines are the model with 
$n_0=2.7\times 10^{-5}~{\rm s^{-1}}$
($\langl T \rangl=2.9$\,hr ), $\beta=0.8$, and the radiation originating
between 1.7$R_*$ and 9$R_*$.  The dashed line is for the
homogeneous model of the same \mdot.  
\label{fig:zoril}} 
\end{figure*} 

The clumped wind model predicts that the  shape of the emission lines
is similar across the spectrum. This prediction  is confirmed
observationally. {\changed \citet{kr03} note from the analysis  of
observed lines that the amount of absorption inferred for different 
lines in \zpup\ and the line shapes are not sensitive to wavelength. 
We emphasize that the model lines shown in 
Fig.\,\ref{fig:pupl1} are not selected best-fit
models. These lines are computed using independently derived 
parameters. Comparing model and observation, we conclude that the
stochastic wind model is capable to reproduce the observed lines 
sufficiently good.}

{\changed Some of the lines shown in
Fig.\,\ref{fig:pupl1}  are contaminated by blends. The
N\,{\sc vii} line is blended with  N\,{\sc vi}\,($\lambda\,24.898$)
\citep{pol06}.  Similarly, Fe\,{\sc xvii} line is possibly  blended
with Fe\,{\sc xix}\,($\lambda\,15.08$). Emission from O\,{\sc vii}
($\lambda\,15.17$)  is also likely to contribute in the far red  wing
of Fe\,{\sc xvii} \citep{osk06}. }

\begin{figure}
\epsfxsize=\columnwidth 
\centering 
\mbox{\epsffile{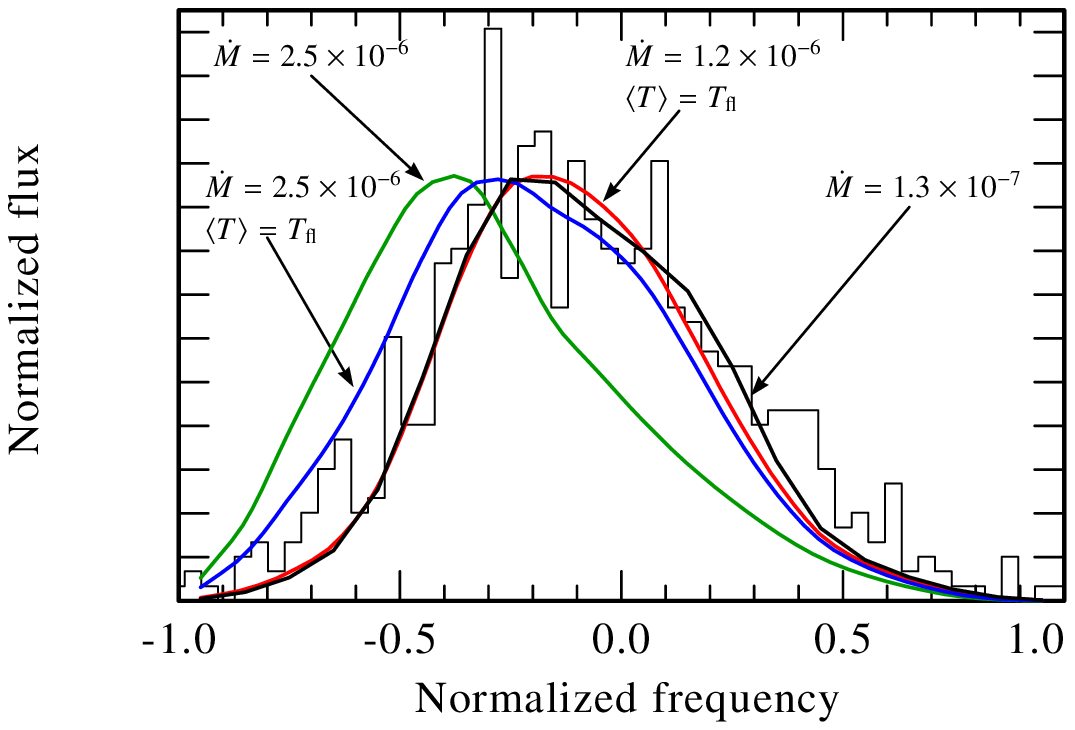}}
\caption{Observed (histogram) and model lines of O\,{\sc viii} in 
\zori. Arrows indicate models used to compute the lines. For smooth wind 
models, only mass-loss rate, $\dot{M}\,[M_\odot\,{\rm yr}^{-1}]$ is shown, 
for clumped wind models, both mass-loss rate and the inverse fragmentation 
frequency, $\langl T \rangl=n_0^{-1}$ is shown. Line with 
$\dot{M}=1.3\times 10^{-7}\,M_\odot\,{\rm yr}^{-1}$ is the best-fit
smooth wind model, as found in \citet{cohen06}.      
\label{fig:difmod}}  
\end{figure} 

\medskip 
\noindent {\bf \xper.} The strongest X-ray emission lines observed in
\xper\  are shown in Fig.\,\ref{fig:xperl}. The data are quite noisy,
however, the  lines look similar to each other regarding broadening and
blue-shifts.  For comparison, we have over-plotted on the  Fe\,{\sc
xvii} and O\,{\sc vii} line both smooth and fragmented model line 
profiles. {\changed The difference between them is  relatively small, 
because the clumps in \xper\  are not optically thick. It appears that
the red wing of the Fe\,{\sc xvii} line  is above the model.  This
could be due to  blending with lines from  higher iron ions, 
or it may indicate that the fragmentation frequency is somewhat 
lower in \xper. \citet{puls06} suggest a much higher clumping 
factor in the outer wind of \xper\ than in \zpup. }
%
\begin{figure*}
\epsfig{figure=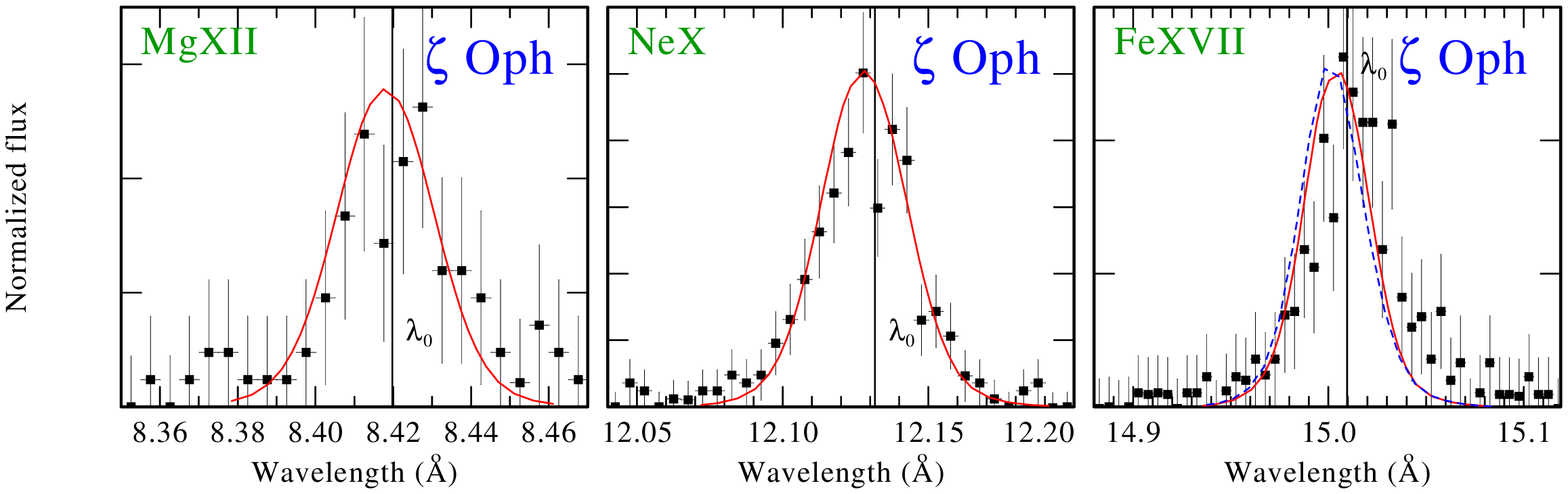, height=18cm, angle=-90} 
\caption{ Lines of Mg\,{\sc xii}, Ne\,{\sc x}, and Fe\,{\sc xvii} observed 
in \zoph\ (co-added {\sc meg}$\pm 1$) and the 
model lines. The solid lines are the model with 
$n_0=2.5\times 10^{-4}~{\rm s^{-1}}$~ ($\langl T \rangl=1.1$\,hr), 
$\beta=1.5$, and the radiation originating
between 1.5\,$R_*$ and 9\,$R_*$.  The wind  optical depth is smaller than 
unity for all these lines. Consequently, all lines can be equally 
 well be reproduced by a
smooth-wind model  (shown by a dashed line). \label{fig:ophl}} 
\end{figure*} 

\begin{figure}
\epsfig{figure=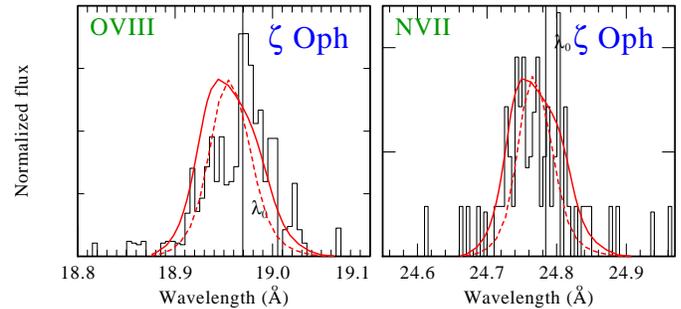, height=8.8cm, angle=-90}
\caption{Lines of O\,{\sc viii} and N\,{\sc vii} 
observed in \zoph\ (histograms, co-added {\sc meg}$\pm 1$) and the
model lines. The solid lines are from a model where radiation 
originates between 1.9\,$R_*$ and 9\,$R_*$, while dashed model lines are for 
the radiation from $1.5\,R_*-9\,R_*$. 
\label{fig:zophON}} 
\end{figure} 

{\changed 
\medskip
\noindent {\bf \zori.} Figure\,\ref{fig:zoril} shows the best-resolved 
lines in the \zori~spectrum. The lines have similar shapes, they are
blue-shifted and broadened. The model lines shown in
Fig.\,\ref{fig:zoril} are calculated using mass-loss which is most
certainly is overestimated by factor of $\sim$few.  Therefore, the
model lines are slightly shifted to the blue compare to the observed
lines. As shown in Fig.\,\ref{fig:linn}, the reduction of fragmentation
frequency by factor of 3 can shift the model line to the red by about
$\approx 0.1\vinf$. The reduction of empirical mass-loss rate of
\zori\ will also lead to the improvement of the model. Lines of \zori\ 
were analyzed by \citet{cohen06} by means of smooth wind model. They
have found from the best-fit model of O\,{\sc viii} that $\kappa_\nu
\dot{M}/4\pi\vinf\approx 0.26$. Substituting the mass-absorption
coefficient from Table\,\ref{tab:kappa} and stellar parameters from
Table\,\ref{tab:wind}, the corresponding mass-loss rate is
$\dot{M}\approx 1.3\times 10^{-7}\,M_\odot\,{\rm yr}^{-1}$, which is
20 times lower than estimated in \citet{lam93}. On the other hand, the
clumped wind model provides the same quality of fit assuming \mdot\ 
only twice as low as in Table\,\ref{tab:wind} and fragmentation
frequency equal inverse flow time. This is illustrated in
Fig.\,\ref{fig:difmod} which shows the observed
and the four model lines of O\,{\sc viii} in \zori. The line from clumped
wind model, with $n_0=T_{\rm fl}^{-1}=9.1\times 10^{-5}\,{\rm s}^{-1}$
and $\dot{M}=1.2\times 10^{-6}\,M_\odot\,{\rm yr}^{-1}$ is nearly
identical with the line computed using smooth wind model with
$\dot{M}\approx 1.3\times 10^{-7}\,M_\odot\,{\rm yr}^{-1}$.}

\medskip
\noindent {\bf \zoph.} The wind of \zoph\ is {\changed almost} transparent 
for X-rays. As can be seen in Fig.\,\ref{fig:tau1}, at the wavelengths
of the Mg\,{\sc xii}, Ne\,{\sc x} and Fe\,{\sc xvii} the wind is
optically thin for the X-rays already above $\approx\,1.5\,R_*$. These
lines are shown in Fig.\,\ref{fig:ophl}. Due to the nearly complete
lack of absorption in the wind the lines provide detailed information
about the location and conditions of the hot plasma itself. The lines
are symmetric and quite narrow, with broadening up to the
$\pm$\,0.8\,\vinf.  The slight asymmetry is due to occultation by the
stellar core. The line shapes indicate that emission originates in the
wind acceleration zone. {\changed The line shape is best reproduced
(by eye estimate) when $\beta=1.5$}.  As can be seen from the two models 
over-plotted on the  observed Fe\,{\sc xvii} line, the difference
between smooth and fragmented models is negligible for the thin wind
of \zoph.

At the wavelengths of the O\,{\sc viii} and the N\,{\sc vii} line the
wind of \zoph\ is thin only above $\approx\,2.5\,R_*$. These lines are
shown in Fig.\,\ref{fig:zophON}. {\changed Both these lines have a low
signal-to-noise ratio, and it is difficult to assess the influence
of wind absorption on line profiles.} The O\,{\sc viii} line appears
asymmetric, with the blue wing being weaker than the red one,
{\changed and our spherically-symmetric model cannot reproduce it
well.  However, it is not clear whether this line asymmetry is
significant. }  The N\,{\sc vii} line is slightly shifted to the
blue.  To demonstrate how sensitive is the emission line profile to
the location of the X-ray emitting plasma, we show two model lines for
O\,{\sc viii} and N\,{\sc vii} in Fig.\,\ref{fig:zophON}. {\changed
For each ion, the model lines differ only by the assumed minimum
radii of X-ray emission.}  It appears that the O\,{\sc viii} and the
N\,{\sc vii} lines originate from slightly further out in the wind
than the lines shown in Fig.\,\ref{fig:ophl} at $\approx 1.9\,R_*$.

The \zoph\ wind opacity is small and X-rays produced in the wind can
escape without suffering significant attenuation. This allows to
assess the intrinsic production of X-rays. We have calculated the
ratios of the X-ray luminosity to the bolometric and mechanical
($\dot{M}v_\infty^2/2$) luminosity of stellar wind in \zoph~and
compared it with these ratios for other stars. The results are given
in Table~\ref{tab:lmech}.
  
The ratio of X-ray to bolometric luminosity in all stars in
Table~\ref{tab:lmech} is slightly smaller than the canonical $10^{-7}$
known for OB stars \citep{berg96}. Perhaps, this is because the large
sample of bright OB stars analyzed by \citet{berg96} is biased towards
binary stars, which tend to be more X-ray bright.

%

\begin{table*} 
  \begin{center}  
   \caption{Ratios of X-ray, bolometric and mechanical luminosities}
\vspace{1em} 
    \renewcommand{\arraystretch}{1.2} 
    \begin{tabular}[h]{lcccccc} 
      \hline \hline
Name&D&$L_{\rm X}$&$L_{\rm bol}$& $L_{\rm mech}$&$L_{\rm X}/L_{\rm bol}$&
$L_{\rm X}/L_{\rm mech}$\\ 
    &[pc]& [erg/s]&[erg/s]& [erg/s]      &               &                  \\
      \hline 
$\zeta$~Pup&460&$2.3\times 10^{32}$&$2.8\times 10^{39}$&$6.7\times 10^{36}$&
$8.2\times 10^{-8}$ & $3.4\times 10^{-5}$ \\ 
$\zeta$~Ori&500&$2.1\times 10^{32}$&$3.0\times 10^{39}$&$3.5\times 10^{36}$&
$6.8\times 10^{-8}$ & $5.8\times 10^{-5}$ \\ 
$\xi$~Per  &850&$1.4\times 10^{32}$&$3.1\times 10^{39}$&$2.3\times 10^{36}$&
$4.5\times 10^{-8}$ & $6.0\times 10^{-5}$\\ 
$\zeta$~Oph&154&$1.2\times 10^{31}$&$2.8\times 10^{38}$&
$\lsim 1.4\times 10^{35}$&
$4.1\times 10^{-8}$ & $\msim 8.5\times 10^{-5}$ \\
\hline \hline 
\noalign{\smallskip}
\multicolumn{7}{l}{Distances are as used in  papers quoted in 
Table~\ref{tab:wind}}\\
\end{tabular} 
    \label{tab:lmech} 
  \end{center} 
\end{table*} 
%
{\changed As can be seen from Table~\ref{tab:lmech} the ratio of the
X-ray and the mechanical luminosity is highest in \zoph\ compared to
the other stars with larger wind attenuation. We estimate that about
0.01\% of the kinetic energy of the stellar wind of \zoph~is spent on
the X-ray generation.  }

\section{Discussion}

{\changed 
The result of our analysis is that the stochastic wind model
lines are in good agreement with the lines observed in stellar X-ray
spectra.  This confirms the clumped nature of O star winds.

So far, modern non-LTE stellar atmosphere models include clumping only
in the first approximation of optically thin clumps. For clumps which
are not optically thin, a non-LTE treatment requires to follow the
transport of radiation inside the clumps, which is presently an
impossible task at least in full detail.

But fortunately, the problem of X-ray formation in the winds of
massive stars is much simpler, since we assume that the X-ray emission
of optically thin plasma is decoupled from the continuum absorption in
cool wind. Therefore, the full non-LTE treatment is not needed, and
the solution of pure-absorption radiative transfer is sufficient.

Optically thin clumping in the winds reduces the empirically
determined mass-loss rates. It is not {\it a priory}
obvious how the presence of optically thick clumps may affect the
mass-loss estimates \citep{brown04}. There is no observational
evidence that the clumps in the wind are {\em all} optically thin. On
the contrary, photometric and spectral variability indicate that
opaque clumps are present in stellar winds. The outstanding
question of ``true`` stellar mass-loss rates in clumped winds is
being extensively investigated at present \citep[e.g.][]{puls06}. 

\subsection{On the dependence of line profiles on mass-loss rate}

In our modeling we adopted mass-loss rates which, when available,
account for the wind clumping.  Since the mass is conserved in the
wind, the optical depth in each clumps scales directly with \mdot, and
inversely with fragmentation frequency (Eq.\,\ref{eq:taujrad}).  The
effect of wind clumping on the lines profiles is most pronounced when
clumps are optically thick for the incident radiation \citep{feld03}.
Therefore, the reduction of empirical mass-loss rates means that in order
for clumps to be optically thick the fragmentation frequency should be
smaller.  
%
%
\begin{figure}
\centering \mbox{\epsfig{figure=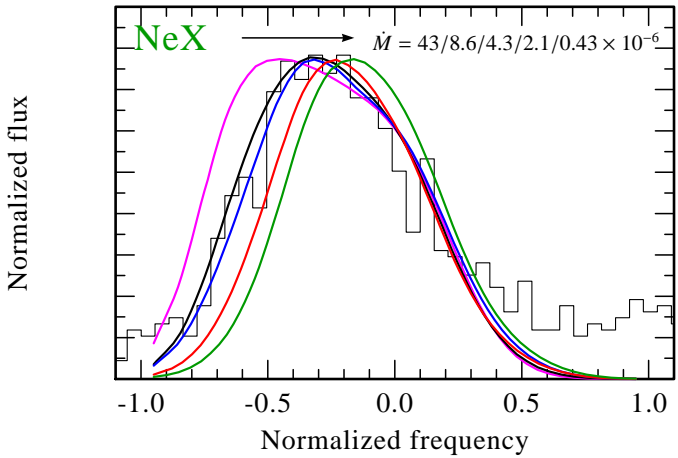,width=6cm,angle=-90}}
\caption{Observed (histogram, co-added {\sc meg}$\pm 1$) and modeled 
Ne\,{\sc x} line in $\zeta$~Pup.  From left to right model line
profiles are calculated \ assuming
$\mdot=43,\,8.6,\,4.3,\,2.1,\,0.43\times 10^{-6}M_\odot$\,yr$^{-1}$.
Corresponding normalization coefficient for the model line maximum to
the observed line maximum is $\eta=2.6\times
10^{4},\,16.0,\,2.6,\,0.85,\,0.23$
\label{fig:pupNemdot}} 
\end{figure} 
%

In Fig.\,\ref{fig:pupNemdot} we model the Ne\,{\sc x} line in \zpup\
for the same fragmentation frequency, but different mass-loss rates.
The lowest value, $\mdot\approx 0.43\times 10^{-6}\,M_\odot\,{\rm
yr}^{-1}$, is an estimate by \citet{ful06}, based on the assumption
that all phosphorus is in the P\,{\sc v} ground level ($q({\rm
P}^{4+})=1$).  Adopting this value, the effect of wind absorption on
the line profile is smaller than observered, as can be seen from the
smaller blueshift of the model line compared to the observed line in
Fig.\,\ref{fig:pupNemdot}. With increasing \mdot\ the line maximum
becomes more and more blueshifted.

The effect of increasing absorption on the {\em line flux}, which
cannot be seen from Fig.\,15 because the profiles are scaled to the
same maximum, is dramatic. Actually, the profile for the highest
\mdot\ is $10^5$ times weaker than for the lowest \mdot, provided
that the same flux has been originally released in the X-ray emitting
zone. Even for the optically thin Fe\,{\sc xvii} line in \zoph\ 
(Fig.\,\ref{fig:ophl}) the flux in the fragmented wind model is about
30\% higher than that in the smooth wind model. Unfortunately, no
models are available yet that would be able to predict the line fluxes
based on detailed hydrodynamic simulations. A comparison of fluxes
between modeled and observed lines would be a very sensitive tool.  }

{\changed 
\subsection{On the influence of the velocity field on the line profile}

The influence of the velocity field on model line profiles is
illustrated in Fig.\,\ref{fig:bet}. Hydrodynamic simulations of
\citet{AF97} indicate that the velocity of the hot plasma is nearly
the same as the cool-wind velocity. In this paper, we approximate the
velocity field as monotonic ( see \citet{feld06} for the treatment of
non-monotonic velocity).  From the width of the line profiles we infer
that the X-ray emission in all four stars is originating in a region
extending from about $\approx 0.3$\vinf\ to $\approx 0.8$\vinf.  The
corresponding radii of emission can be obtained if the velocity-law
parameter $\beta$ is known. Vice versa, if radii of emission are
given, $\beta$ can be inferred from the line-profiles.  Usually,
$\beta$ is estimated from UV line fits, and typically lies between 0.5
and 1.5. \citet{ful06} obtained for \zpup\ $\beta=0.5$, while
\citet{puls06} obtain  $\beta=0.95$ and notice
that different values may apply for inner and outer parts of the wind.
$\beta\approx 1$ was inferred from the line profile variability in
\zpup\ \citep{ev98}. It seems that optical and UV line profiles are
less sensitive to the exact value of $\beta$ than the X-ray lines
studied here. As shown in Fig.\,\ref{fig:bet}, the line profile changes
significantly in dependence on $\beta$. The synthetic lines for our
four program stars shown in the figures were actually slightly
optimized by adjusting $\beta$ within the uncertainties of previous
estimates (see figure captions).}
   
\begin{figure}
\centering \mbox{\epsfig{figure=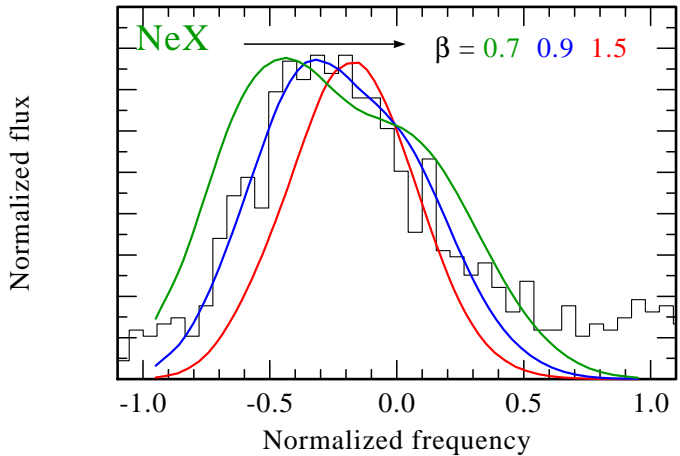,width=6cm,angle=-90}}
\caption{Observed (histogram, co-added {\sc meg}$\pm 1$) and modeled 
Ne\,{\sc x} line in $\zeta$~Pup. From left
to right line profiles are calculated assuming
$\beta=0.7,\,0.9,\,1.5$.
\label{fig:bet}} 
\end{figure} 

\subsection{On the applicability of smooth wind radiative transfer 
in clumped winds} 

The optical depth in a smooth wind is given by Eq.\,(\ref{eq:tauhz}),
which was employed by \citet{kr03} and \citet{cohen06} to fit the line
profiles observed in the X-ray spectra of \zpup\ and \zori.  They have
used $\tau_0=\kappa_\nu\mdot/4\pi v_\infty$ as one of their four free
fitting parameters. Assuming $\kappa_\nu=125$\,g\,cm$^{-2}$ for all
analyzed lines in \zori, it was shown that the best-fitting values of
$\tau_0$ imply significantly reduced mass-loss rates, e.g. due to the 
wind clumping.  In a clumped wind, however, the optical
depth is given by Eq.\,(\ref{eq:tint}), which is more general than
Eq.\,(\ref{eq:tauhz}). The latter applies only for a limiting case of
numerous and {\it optically thin} clumps.

Little is known so far about mass, density and geometry of individual
clumps. Clump optical depths are likely to be different at different
radii in the wind. This is reflected in our stochastic wind model,
where the mass of a clump is conserved, but its radial optical depth
scales as $r^{-2}$ (Eq.\,\ref{eq:taujrad}). Furthermore, since clump
optical depth is proportional to $\kappa_\nu$, a clump can be
optically thick at one wavelength and optically thin at another. As
can be seen from Table\,\ref{tab:kappa}, the mass absorption
coefficient differs by one order of magnitude between the wavelenghts
of the Si\,{\sc xiv} and the O\,{\sc viii} line.
Figure\,\ref{fig:pupl1} shows model lines
of Si\,{\sc xiv} and O\,{\sc viii} in \zpup. For the Si\,{\sc xiv}
line the difference between clumped and smooth model is small,
indicating that the clumps are optically thin at
$\lambda\,6.18$\,(Si\,{\sc xiv}).  On the other hand, the model lines
of O\,{\sc viii} for smooth and clumped wind are significantly
different, and, therefore, at least some fraction of the clumps are
optically thick at $\lambda\,18.97$\,(O\,{\sc viii}). Consequently,
Eq.\,(\ref{eq:tint}) can be applied for the Si\,{\sc xiv} line, but 
cannot be used  for the O\,{\sc viii} line.

Overall, when stellar mass-loss rates are revised down because of the
wind clumping, the clumped wind radiative transfer should be applied
for consistency.  }

\subsection{On the line profiles and clump geometry}

{\changed 
As discussed in detail in \citet{osk04}, the optical depth in a clumped
wind is determined by two factors.  The first factor is the
probability that a photon which encounters a clump is absorbed by
clump material. This probability is obviously equal to one for
optically thick clumps, $\tau_{\rm cl} \gg 1$, and is reduced
otherwise to {\mbox{$1-\exp(-\tau_{\rm cl})$}} (see Eq.\,\ref{eq:tint}).
 
The second factor is the probability that a photon propagating along
the ray towards the observer does encounter a clump. When clumps are
distributed randomly in the wind, this probability depends on the
length of the path. The longer a photon travels through the wind,
the higher is the probability that it will be intercepted by a clump.
Therefore, photons emitted in the part of the wind which is receding
from the observer have higher probability to be absorbed. 
Thus there are less photons in the red wing of a line reaching
an observer than in the blue part. This is the basic physical reason
for the blueshift of the X-ray emission line profiles. 

Moreover, the probability that a photon does encounter a clump depends
also on the clump cross-sections. Approximating clumps as spheres, the
clump cross-section is the same from all directions, which is conceptually 
similar to the atomic cross-sections. The emergent line profiles are 
skewed in both cases.

\begin{figure}
\epsfxsize=\columnwidth 
\centering \mbox{\epsffile{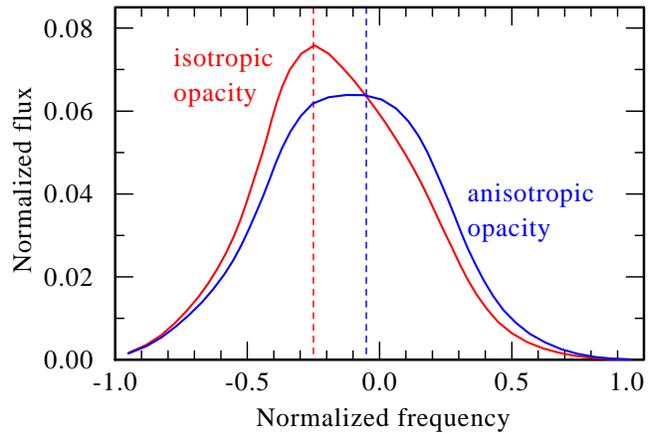}}
\caption{Two model lines, both calculated with  
$n_0=4.1\times 10^{-5}~{\rm s^{-1}}$ and $\tau_0=10$. 
Flux is normalized to the unabsorbed line flux. The only difference
between the models is the assumption of isotropic opacity (i.e.\ clumps as
``balls'') and anisotropic opacity (i.e.\ clumps as ``pancakes'').
The solid vertical lines mark the positions of line maxima. 
Isotropic opacity leads to a more blueshifted and significantly skewed 
line profile.
\label{fig:shape}} 
\end{figure} 

To explain the observed near-symmetry of line profiles in the
framework of the smooth wind model, \citet{kr03} and \citet{cohen06}
suggested the reduction of mass-loss rates, which implies lower
densities. When the particle number density is reduced, the photon
mean free path is increased. In a clumped wind with spherical clumps,
each clump is analogous to a particle. In order for the line profile
to be symmetric, the ``particle'' (i.e.\ clump) number density must be
reduced, which consequently leads to a larger photon mean free path
(i.e.\ porosity length).

This deep analogy between smooth winds on one side and clumped winds
with spherical clumps on the other side explains the principal similarity
between the conclusions reached in \citet{cohen06} and in \citet{ow06},
namely the requirement of large mean free paths in order to produce
symmetric emission line profiles. 

A qualitatively different behavior of the radiative transfer is
achieved, however, for anisotropic opacity (i.e.\ non-spherical,
oriented clumps), as studied in \citet{feld03} and \citet{osk04}.
This effect of clump geometry on emergent line profiles is
demonstrated in Fig.\,\ref{fig:shape}. Two model lines are calculated
with the the same fragmentation frequency and the same value of
$\tau_0$.  The only difference is that in one case isotropic opacity,
i.e. spherical ``ball-like'' clumps are assumed, while the other model
is calculated for the anisotropic opacity in the limiting case of
clumps which are flat like ``pancakes''. Figure\,\ref{fig:shape}
clearly shows that in the latter case the line is non-skewed and less
blueshifted than in the former case.
  
The assumption of flat clumps was used in the present paper to model
the observed line profiles. We have demonstrated that this model
yields good agreement with the observed data.}

\section{Conclusions}

The effect of clumping on the wind attenuation is threefold.  Firstly,
it reduces the wind absorption compared to the smooth-wind model,
allowing for more flux to escape. Secondly, it leads to a weaker
dependence of the effective opacity on wavelength, making it entirely
grey in the case of optically thick clumps. This prediction of similar
opacity for different lines agrees well with the similarity of line
profiles observed at different wavelengths. Thirdly, it affects the
shape of line profiles, {\changed but only in the case when effective
anisotropic opacity is assumed.}  The radially compressed, slab-like 
shell fragments lead to the emergence of the symmetric, but blue-shifted
lines profiles which are indeed observed.

Inhomogeneity of the stellar wind can reduce the mass-loss rates
estimated empirically from the analysis of recombination and H$\alpha$
lines.  {\changed These reduced mass-loss rates should be used in
conjunction with the stochastic wind radiative transfer formalism in
order to consistently model X-ray emission line profiles.}

The radii in the wind where continuum optical depth for the X-rays is
unity is below 10\,$R_*$ in Chandra MEG energy band for
\zpup,\,\zori,\,\xper, and \zoph.  Stellar wind in \zoph\ is almost 
transparent for the X-rays, consequently the emission line profiles
show no effect of wind absorption. We estimate that $\sim\,0.01$\% of
the mechanical energy of the wind of \zoph\ is converted into X-ray
emission. The X-ray emission lines observed in the spectra of
\zpup,\,\zori,\, and
\xper\ are affected by wind absorption. This result, as well as shape
of line profiles, is in agreement with the analysis of line ratios in
the He-like ions. The above analysis constrains the origin of X-ray
emission to the acceleration zone in the stellar wind.

{\changed Summarizing, we conclude that by using independently
determined parameters as input for the model of an inhomogeneous
stellar wind } we are capable to reproduce the observed X-ray emission
lines in O stars. The wind clumping explains the shape of the line
{\em profiles}, i.e.\ their width, symmetry and blueshift, as well as their 
similarity across the spectrum.

\section*{Acknowledgments}
The observational data are obtained using Chandra Data Archive.  This
research has made use of the SIMBAD database, operated at CDS,
Strasbourg, France, and of data obtained through the High Energy
Astrophysics Science Archive Research Center Online Service, provided
by the NASA/Goddard Space Flight Center. The paper has benefited from
useful discussions with W. Waldron, S.P. Owocki, and
J.P. Cassinelli. We also appreciate comments made by D. Cohen and an
anonymous referee which led to the improvements of the paper. LMO and
AF acknowledge support from Deutsche Forschungsgemeinschaft grants Fe
573/1-1 and Fe 573/3-P. 

{}

\label{lastpage}


\begin{thebibliography}{}

\bibitem[\protect\citeauthoryear{Bergh{\"o}fer et al.}{1996}]{berg96}
Bergh{\"o}fer T.W., Schmitt J.H.M.M., Cassinelli J.P., 1996, A\&AS, 118, 481

\bibitem[\protect\citeauthoryear{Bouret et al.}{2005}]{bou05}
Bouret J.-C., Lanz T., Hillier D.J., 2005, A\&A, 438, 301

\bibitem[\protect\citeauthoryear{Brown et al.}{2004}]{brown04}
Brown J.C., Cassinelli J.P., Li Q., Kholtygin A.F ., Ignace R. 2004, 
A\&A, 426, 323

\bibitem[\protect\citeauthoryear{Cassinelli \& Olson}{1979}]{cas79}
Cassinelli, J.P. \& Olson, G.L. 1979, ApJ, 229, 304

\bibitem[\protect\citeauthoryear{Cassinelli et al.}{2001}]{cas01}
Cassinelli J.P., Miller N.A., Waldron W.L., MacFarlane J.J., Cohen D.H.,
2001, ApJ, 554, L55

\bibitem[\protect\citeauthoryear{Castor, Abbot \& Klein}{1975}]{CAK75} 
Castor J.I., Abbott D.C., Klein R.I., 1975, ApJ, 195, 157

\bibitem[\protect\citeauthoryear{Cohen et al.}{2006}]{cohen06}
Cohen D.H., Leutenegger M.A., Grizzard K.T., Reed C.L., Kramer R.H., 
Owocki S.P, 2006, MNRAS, 368, 1905  

\bibitem[\protect\citeauthoryear{Dessart \& Owocki}{2003}]{des03}
Dessart L. \& Owocki S.P., 2003, A\&A, 406, L1 

\bibitem[\protect\citeauthoryear{Eversberg et al.}{1998}]{ev98}
Eversberg T., L{\`e}pine S., Moffat A.F.J., 1998, ApJ, 494,799

\bibitem[\protect\citeauthoryear{Feldmeier et al.}{1997}]{AF97} 
Feldmeier A., Puls J., Pauldrach A.W.A., 1997, A\&A, 322, 878

\bibitem[\protect\citeauthoryear{Feldmeier et~al.}{2003}]{feld03} 
Feldmeier A., Oskinova L., Hamann W.-R., 2003, A\&A, 403, 217

\bibitem[\protect\citeauthoryear{Feldmeier \& Nikutta}{2006}]{feld06}
Feldmeier,A. \& Nikutta R., 2006, A\&A, 446, 661

\bibitem[\protect\citeauthoryear{Fullerton et al.}{2006}]{ful06}
Fullerton A.W., Massa D.L., Prinja R.K., 2006, ApJ, 637, 1025

\bibitem[\protect\citeauthoryear{Gabriel \& Jordan}{1969}]{gab69}
Gabriel A.H., \& Jordan C., 1969, MNRAS, 145, 241 

\bibitem[\protect\citeauthoryear{Ignace}{2001}]{ig01}
Ignace R., 2001, ApJ, 549, L119

\bibitem[\protect\citeauthoryear{Hamann \& Koesterke}{1998}]{hk98}
Hamann W.-R. \& Koesterke L., 1998, A\&A, 335, 1003

\bibitem[\protect\citeauthoryear{Hamann \& Gr{\"a}fener}{2004}]{powr}
Hamann W.-R. \& Gr{\"a}fener G., 2004, A\&A, 427, 697 

\bibitem[\protect\citeauthoryear{Harnden et al.}{1979}]{har79}
Harnden F.R., Jr., et al., 1979, ApJ, 234, L51

\bibitem[\protect\citeauthoryear{Henley et al.}{2003}]{Henley03}
Henley D.B., Stevens I.R., Pittard J.M., 2003, MNRAS, 346, 773

\bibitem[\protect\citeauthoryear{Hillier et al.}{1993}]{hil93}
Hillier D.J., Kudritzki R.P., Pauldrach A.W., Baade D., Cassinelli J.P., 
Puls J., Schmitt J.H.M.M., 1993, A\&A 276, 117

\bibitem[\protect\citeauthoryear{Hoogerwerf et al.}{2001}]{Hoog01}
Hoogerwerf R., de Bruijne J.H.J., Zeeuw P.T., 2001, A\&A, 365, 49

\bibitem[\protect\citeauthoryear{Hummel et al.}{2000}]{hum00}
Hummel C.A., White N.M., Elias II N.M., Hajian A.R., Nordgren T.E., 
2000, ApJ, 540, L91  

\bibitem[\protect\citeauthoryear{Kahn et al.}{2001}]{kahn01}
Kahn S.M., Leutenegger M.A., Cotam J., Rauw G., Vreux J.-M., 
den Boggende A.J.F., Mewe R., G{\"u}del M., 2001, A\&A, 276, 117 

\bibitem[\protect\citeauthoryear{Kramer et al.}{2003}]{kr03}
Kramer R.H., Cohen D.H., Owocki S.P., 2003, ApJ, 592, 532

\bibitem[\protect\citeauthoryear{Lamers \& Leitherer}{1993}]{lam93}
Lamers H.J.G.L.M. \& Leitherer C, 1993, ApJ, 412, 771

\bibitem[\protect\citeauthoryear{Lamers et al.}{1999}]{lam99}
Lamers H.J.G.L.M., Haser S., de Koter A., Leitherer C., 1999, ApJ, 516, 872

\bibitem[\protect\citeauthoryear{Lucy \& Solomon}{1970}]{Lucy70}
Lucy L.B. \& Solomon P.M., 1970, ApJ, 159, 879

\bibitem[\protect\citeauthoryear{MacFarlane et al.}{1991}]{mcf91}
MacFarlane J.J., Cassinelli J.P., Welsh B.Y., Vedder P.W., 
Vallerga J.V., Waldron W.L., 1991, ApJ, 380, 564

\bibitem[\protect\citeauthoryear{Markova et al.}{2005}]{mar05}
Markova N., Puls J., Scuderi S., Markov H., 2005, A\&A, 440, 1133

\bibitem[\protect\citeauthoryear{Miller et al.}{2002}]{mil02}
Miller N.A., Cassinelli J.P., Waldron W.L., MacFarlane J.J., Cohen D.H.,
2002, ApJ, 577, 951

\bibitem[\protect\citeauthoryear{Oskinova et al.}{2004}]{osk04}
Oskinova L.M., Feldmeier A., Hamann W.-R., 2004 A\&A 422, 675

\bibitem[\protect\citeauthoryear{Oskinova et al.}{2006}]{osk06}
Oskinova L.M., Hamann W.-R., Feldmeier A., 2006, in Proc.
``High resolution X-ray spectroscopy: towards XEUS and Con-X'',  
astro-ph/0605560

\bibitem[\protect\citeauthoryear{Owocki \& Rybicki}{1984}]{OR84}
Owocki S.P. \& Rybicki G.B., 1984, ApJ, 284, 337

\bibitem[\protect\citeauthoryear{Owocki \& Cohen}{2006}]{ow06}
Owocki S.P. \& Cohen D.H., 2006, ApJ, in press (astro-ph/0602054) 

\bibitem[\protect\citeauthoryear{Paerels \& Kahn}{2003}]{paer03}
Paerels F.B.S. \& Kahn S.M., 2003, ARA\&A, 41, 291

\bibitem[\protect\citeauthoryear{Pollock et al.}{2005}]{pol05}
Pollock A.M.T., Corcoran M.F., Stevens I.R., Williams P.M., 
2005, ApJ, 629, 482

\bibitem[\protect\citeauthoryear{Pollock \& Raassen}{2006}]{pol06}
Pollock A.M.T. \& Raassen A., 2006, A\&A , submitted

\bibitem[\protect\citeauthoryear{Pomraning}{1991}]{pom91}
Pomraning G.C., 1991, Linear kinetic theory and particle transport in 
stochastic mixtures, Singapore; New Jersey: World Scientific. Series on 
advances in mathematics for applied sciences 7

\bibitem[\protect\citeauthoryear{Puls et al.}{2006}]{puls06}
Puls J., Markova N., Scuderi S., Stanghellini C., Taranova O.G., 
Burnley A.W., Howarth I.D., 2006, A\&A, 454, 625

\bibitem[\protect\citeauthoryear{Porquet et~al.}{2001}]{por01}
Porquet D., Mewe R., Dubau J., Raassen A.J.J., Kaastra J.S., 2001, A\&A, 
376, 1113

\bibitem[\protect\citeauthoryear{Repolust et~al.}{2004}]{Rep04}
Repolust T., Puls J., Herrero A., 2004, A\&A, 415, 349

\bibitem[\protect\citeauthoryear{Runacres \& Owocki} {2005}]{RO05}
Runacres M.C. \& Owocki S.P., 2005, A\&A, 429, 323

\bibitem[\protect\citeauthoryear{Sako et al.}{2003}]{sako03}
Sako M., Kahn S.M., Paerels F., Liedahl D.A., Watanabe S., Nagase F., 
Takahashi T., 2003, in High Resolution X-ray Spectroscopy with 
{\em XMM-Newton} and {\em Chandra}, ed. G. Branduari-Raymont 
(astro-ph/p309503)  

\bibitem[\protect\citeauthoryear{Schild et al.}{2004}]{sch04}
Schild H., et al., 2004, A\&A, 422, 177 

\bibitem[\protect\citeauthoryear{Seward et al.}{1979}]{sew79}
Seward F.D., Forman W.R., Giacconi R., Griffiths R.E., Harnden F.R., 
Jr., Jones C., Pye J.P., 1979, ApJ, 234, L55

\bibitem[\protect\citeauthoryear{Van der Meer et al.}{2005}]{vdm05}
van~der~Meer A., Kaper L., Di~Salvo T., et al., 2005, A\&A, 432, 999

\bibitem[\protect\citeauthoryear{Waldron \& Cassinelli}{2001}]{wal01}
Waldron W.L. \& Cassinelli J.P., 2001, ApJ, 548, L45

\bibitem[\protect\citeauthoryear{Waldron}{2005}]{wal05}
Waldron W.L. 2005, in Proc. ``The Nature and Evolution of Disks Around 
Hot Stars``
ASP Conference Series, Vol. 337, Ed. R. Ignace \& K. G. Gayley, 329

\bibitem[\protect\citeauthoryear{Wojdowski \& Schulz}{2005}]{woj05}
Wojdowski P.S. \& Schulz, N.S., 2005, ApJ, 627, 953

\end{thebibliography}
\end{document}